\documentclass[aps,prl,twocolumn,preprintnumbers,amsmath,amssymb,superscriptaddress]{revtex4-2}
\usepackage{amsmath,graphicx}
\usepackage[utf8]{inputenc}
\usepackage[T1]{fontenc}
\usepackage{xcolor}
\usepackage{textcomp}
\usepackage{bm}

\usepackage{siunitx}
\usepackage{physics}
\usepackage{amsmath}
\usepackage{tikz}
\usepackage{mathdots}
\usepackage{yhmath}
\usepackage{cancel}
\usepackage{color}
\usepackage{array}
\usepackage{multirow}
\usepackage{amssymb}
\usepackage{gensymb}
\usepackage{tabularx}
\usepackage{extarrows}
\usepackage{booktabs}
\usetikzlibrary{fadings}
\usetikzlibrary{patterns}
\usetikzlibrary{shadows.blur}
\usetikzlibrary{shapes}

\usepackage{color}
\definecolor{LinkColor}{rgb}{0.75,0.0,0.2}

\usepackage{hyperref}
\hypersetup{
	pdfauthor={good guys},
	pdftitle={good title},
	colorlinks=true,
	citecolor=LinkColor,
	linkcolor=LinkColor,
	urlcolor=LinkColor,
}

\usepackage{listings}
\definecolor{lightgray}{gray}{1}

\usepackage{theorem}

\usepackage{tikz}

\newcommand{\nc}{\newcommand}
\nc{\braoprket}[3]{\langle#1|#2|#3\rangle}
\nc{\opn}[1]{\operatorname{#1}}
\nc{\avg}[1]{\langle#1\rangle}
\nc{\ketbrasame}[1]{|#1\rangle\!\langle#1|}
\nc{\swap}{\opn{SWAP}}
\nc{\E}{\mathbb{E}}
\nc{\Var}{\opn{Var}}
\nc{\dg}{\dagger}

\usepackage[normalem]{ulem}
\nc{\hknew}[1]{\textcolor{brown}{#1}}

\begin{document}
\title{A Unified Variational Framework for Quantum Excited States}

\author{Shi-Xin Zhang}
\email{shixinzhang@iphy.ac.cn}
\affiliation{Institute of Physics, Chinese Academy of Sciences, Beijing 100190, China}
\author{Lei Wang}
\email{wanglei@iphy.ac.cn}
\affiliation{Institute of Physics, Chinese Academy of Sciences, Beijing 100190, China}

\date{\today}

\begin{abstract}
Determining quantum excited states is crucial across physics and chemistry but presents significant challenges for variational methods, primarily due to the need to enforce orthogonality to lower-energy states, often requiring state-specific optimization, penalty terms, or specialized ansatz constructions. We introduce a novel variational principle that overcomes these limitations, enabling the \textit{simultaneous} determination of multiple low-energy excited states. The principle is based on minimizing the trace of the inverse overlap matrix multiplied by the Hamiltonian matrix, $\Tr(\mathbf{S}^{-1}\mathbf{H})$, constructed from a set of \textit{non-orthogonal} variational states $\{|\psi_i\rangle\}$. Here, $\mathbf{H}_{ij} = \langle\psi_i | H | \psi_j\rangle$ and $\mathbf{S}_{ij} = \langle\psi_i | \psi_j\rangle$ are the elements of the Hamiltonian and overlap matrices, respectively. This approach variationally optimizes the entire low-energy subspace spanned by $\{|\psi_i\rangle\}$ without explicit orthogonality constraints or penalty functions. We demonstrate the power and generality of this method across diverse physical systems and variational ansatzes: calculating the low-energy spectrum of 1D Heisenberg spin chains using matrix product states, finding vibrational spectrum of Morse potential using quantics tensor trains for real-space wavefunctions, and determining excited states for 2D fermionic Hubbard model with variational quantum circuits. In all applications, the method accurately and simultaneously obtains multiple lowest-lying energy levels and their corresponding states, showcasing its potential as a unified and flexible framework for calculating excited states on both classical and quantum computational platforms.
\end{abstract}

\maketitle

\textit{Introduction.---} 
Quantum excited states are fundamental to understanding a vast range of physical and chemical phenomena, from the electronic and vibrational spectroscopy of molecules \cite{Huber1979} to the properties of quantum many-body systems \cite{Wlfle2018, Pal2010a, Chen2024}. Calculating these states accurately remains a significant challenge for theoretical and computational methods. The standard variational principle, based on minimizing the expectation value $\langle\psi|H|\psi\rangle$, is a powerful tool for finding the ground state, but its direct application to excited states is complicated because excited states minima are subject to the constraint of being orthogonal to all lower-energy eigenstates.

Existing variational approaches typically manage this orthogonality constraint through state-specific optimizations within fixed sectors of given symmetry or momentum~\cite{stlund1995,Pirvu2012,Haegeman2012,Vanderstraeten2018,Zou2018, Vanderstraeten2019,Ponsioen2020, Tu2021}, sequentially targeting states while enforcing orthogonality to previously found lower-energy solutions \cite{White1992, Bauls2013, Choo2018, Jones2019, Pathak2021, Entwistle2023, Larsson2025}, adding penalty terms to the loss function that penalize non-orthogonality between a set of states \cite{Higgott2019, Wheeler2024, Quiroga2025}, or utilizing specialized ansatz constructions that inherently orthogonal with potentially limited expressiveness \cite{Nakanishi2019, LaRose2019, Parrish2019, Li2024, zhang2024neural}. While successful in many cases, these methods often face limitations such as sensitivity to initialization, numerical instability associated with penalty terms, the need for costly sequential calculations, or restrictions on the functional form of the variational states. 
There are other approaches targeting eigenstates directly by minimizing the energy variance $\langle H^2 \rangle - \langle H \rangle^2$ \cite{Umrigar1988, Siringo2005, Umrigar2005, Pollmann2016, Vicentini2019, Zhang2021, Liu2021c,Zhang2022} or the squared difference from a target energy $(H-\lambda)^2$ \cite{Wang1994, McClean2016, Santagati2018, Wang2023, Cenedese2024}. While these methods can find eigenstates, the obtained states are generally not the desired low-lying excited states.

Here, we introduce a novel variational principle for the simultaneous determination of multiple low-energy quantum excited states that circumvents the need for explicit orthogonality constraints or penalty terms during the optimization. Our approach minimizes the trace of the inverse overlap matrix multiplied by the Hamiltonian matrix, $L = \Tr(\mathbf{S}^{-1}\mathbf{H})$, constructed from a set of $N_s$ non-orthogonal variational states $\{|\psi_i\rangle\}$. This loss function directly targets multiple lowest eigenvalues within the subspace spanned by the variational states. The loss function in this form, explored in the context of machine learning~\cite{Edelman1998, Pfau2019}, is rarely employed to tackle quantum many-body problems \cite{Pfau2024}. We demonstrate the generality and effectiveness of this new variational framework by applying it to diverse problems and variational ansatzes: the low-energy spectrum of 1D spin chains using matrix product states (MPS)~\cite{Schollwock2011,Stoudenmire2012}, vibrational excitations of a Morse potential with quantics tensor train (TT) wavefunctions \cite{Khoromskij2011, NezFernndez2025, Lubasch2018, Garca-Ripoll2021, Ye2022, Gourianov2022, Shinaoka2023, Ritter2024}, and excited states of 2D fermionic Hubbard model with variational quantum circuits \cite{Peruzzo2014, Bharti2021, Cerezo2020b, Tilly2022}. Our results show accurate and simultaneous convergence to multiple lowest-lying energy states across these applications, highlighting the potential of this approach as a flexible and unified framework for variational excited state calculations.

\begin{figure}[t]\centering
	\includegraphics[width=0.5\textwidth]{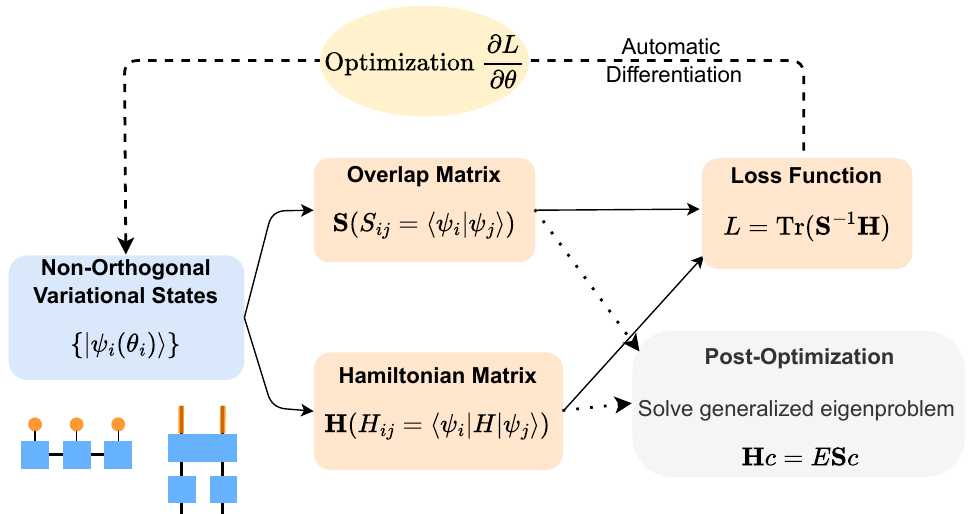}
	\caption{End-to-end variational optimization with the loss function $L = \mathrm{Tr}(\mathbf{S}^{-1}\mathbf{H})$ for simultaneously obtaining multiple low-lying excited states. }
\label{fig:workflow}
\end{figure}

\textit{Method.---} 
Our method seeks to variationally determine a set of $N_s$ lowest energy eigenstates for quantum many-body systems simultaneously. Crucially, we do not impose any explicit orthogonality constraints on these variational states during the optimization.
We begin with a set of $N_s$ variational quantum many-body states $|\psi_i(\vec{\theta}_i)\rangle$, where $\vec{\theta}_i$ represents the variational parameters for state $i$. The total parameter set for the optimization is $\vec{\theta} = \{\vec{\theta}_1, \dots, \vec{\theta}_{N_s}\}$. These states need not be normalized or mutually orthogonal. For a given set of parameters $\vec{\theta}$, two $N_s \times N_s$ matrices: the overlap matrix $\mathbf{S}$ and the Hamiltonian matrix $\mathbf{H}$ are constructed as:
\begin{align}
\mathbf{S}_{ij}(\vec{\theta}) &= \langle\psi_i(\vec{\theta}_i) | \psi_j(\vec{\theta}_j)\rangle ,\\
\mathbf{H}_{ij}(\vec{\theta}) &= \langle\psi_i(\vec{\theta}_i) | H | \psi_j(\vec{\theta}_j)\rangle,
\end{align}
where $H$ is the system Hamiltonian. Assuming the variational states $\{|\psi_i\rangle\}$ are linearly independent, the overlap matrix $\mathbf{S}$ is Hermitian and positive definite, thus invertible. 

The core of our method is the minimization of the following loss function with respect to the variational parameters $\vec{\theta}$:
\begin{equation}
L(\vec{\theta}) = \Tr(\mathbf{S}(\vec{\theta})^{-1}\mathbf{H}(\vec{\theta})). \label{eq:loss}
\end{equation}
The theoretical motivation for minimizing this quantity lies in its connection to the generalized eigenvalue problem and the minimax principle for Ritz values. The loss is equivalent to the sum of eigenvalues, i.e., $L = \Tr(\mathbf{S}^{-1}\mathbf{H}) = \sum_{\alpha=1}^{N_s} E_\alpha$, where $E_\alpha$ are the approximate energies obtained from the current set of variational states $\{|\psi_i\rangle\}$. For any set of $N_s$ linearly independent basis states $\{|\psi_i\rangle\}$ spanning a finite-dimensional subspace, the eigenvalues $E_\alpha$ and expansion coefficients $c_\alpha$ for the eigenstates $|\Psi_\alpha\rangle = \sum_{i=1}^{N_s} c_{i\alpha} |\psi_i\rangle$ within this subspace are given by the generalized eigenvalue problem $\mathbf{H}c = E \mathbf{S}c$.
These eigenvalues, $E_1 \le E_2 \le \dots \le E_{N_s}$, are known as Ritz values. According to the minimax principle, the Ritz values obtained from any $N_s$-dimensional subspace are always upper bounds to the corresponding exact eigenvalues of the full Hamiltonian $E_1^{exact} \le E_2^{exact} \le \dots \le E_{N_s}^{exact}$ \footnote{See supplemental materials for details}:
\begin{equation}
E_\alpha(\vec{\theta}) \ge E_\alpha^{exact} \quad \text{for } \alpha = 1, \dots, N_s.
\end{equation}

By minimizing this sum, the optimization drives the variational states $\{|\psi_i\rangle\}$ to span the subspace corresponding to the lowest possible sum of $N_s$ energy levels, which approximate the lowest $N_s$ energy levels of the full Hamiltonian if the ansatz is sufficiently expressive.
We minimize the loss function $L(\vec{\theta})$ using standard gradient-based optimization techniques. The gradient $\nabla_{\vec{\theta}} L$ is computed efficiently via automatic differentiation frameworks \cite{Liao2019, Zhang2019b}. 
After the optimization converges and the parameters $\vec{\theta}$ are optimized, the final approximate low-energy excited states and their energies are obtained by solving the generalized eigenvalue problem $\mathbf{H}c = E \mathbf{S}c$ using optimized $\mathbf{H}$ and $\mathbf{S}$ matrices. The resulting eigenvalues $E_\alpha$ are the approximate energies, and the eigenvectors $c_\alpha$ provide the coefficients to construct the final orthonormal approximate eigenstates $|\Psi_\alpha\rangle = \sum_{i=1}^{N_s} c_{i\alpha} |\psi_i\rangle$. The end-to-end workflow of our proposed variational principle for excited states calculation is shown in Fig.~\ref{fig:workflow}.

To demonstrate the generality and effectiveness of the proposed variational framework, we apply it to three distinct quantum many-body problems employing different classes of variational ansatzes. The numerical simulation is implemented by quantum software {\sf TensorCircuit-NG}~\cite{*[{ }] [{. https://github.com/tensorcircuit/tensorcircuit-ng.}] Zhang2022_z}.

\textit{Results for 1D spin chain with periodic matrix product states.---}
\label{sec:heisenberg_mps}
We first apply our variational method to calculate the low-energy spectrum of a one-dimensional spin-half chain with $N=16$ sites and periodic boundary conditions (PBC). The Hamiltonian is given by:
\begin{equation}
H = \sum_{i=1}^{L} \frac{1}{4} (J_x \sigma_i^x \sigma_{i+1}^x + J_y \sigma_i^y \sigma_{i+1}^y + J_z \sigma_i^z \sigma_{i+1}^z) + h_z \sigma_i^z,\label{eq:heis}
\end{equation}
where $\sigma_i^{x,y,z}$ are the Pauli matrices at site $i$. This model serves as a general benchmark for methods dealing with low-energy spectrum of many-body systems.

We employ a set of $N_s=32$ independent MPS with PBC to represent the variational states $\{|\psi_i\rangle\}$. Each MPS $|\psi_k(\{\mathbf{A}_k^i\})\rangle$ is parameterized by a collection of rank-3 tensors $\mathbf{A}_k^i$ at each site $i$, with bond dimension $\chi=16$. The wavefunction amplitude is given by $\Psi_{s_1 s_2 \dots s_N} = \mathrm{Tr} \left(\mathbf{A}_1(s_1) \mathbf{A}_2(s_2) \dots \mathbf{A}_N(s_{N})\right)$. We don't need to normalize the state within the ansatz as the overall normalization and overlaps $\langle \psi_i|\psi_j\rangle$ are explicitly handled as $\mathbf{S}$. All $2N_sN\chi^2$ trainable parameters are randomly initialized from the standard normal distribution and variationally optimized.

The calculation of the overlap matrix elements $\mathbf{S}_{ij} = \langle\psi_i | \psi_j\rangle$ and Hamiltonian matrix elements $\mathbf{H}_{ij} = \langle\psi_i | H | \psi_j\rangle$ for periodic MPS involves efficient tensor network contractions, scaling polynomially with system size $L$ and bond dimension $\chi$, specifically, $O(N_s^2 N \chi^6)$ for matrix elements in the standard transfer-matrix approach. The gradients of the loss function $L = \Tr(\mathbf{S}^{-1}\mathbf{H})$ with respect to the MPS tensors are computed using efficient backpropagation algorithms adapted for tensor networks \cite{Liao2019}. We minimize $L$ using a gradient-based optimizer L-BFGS. 

\begin{figure}[t]\centering
	\includegraphics[width=0.43\textwidth]{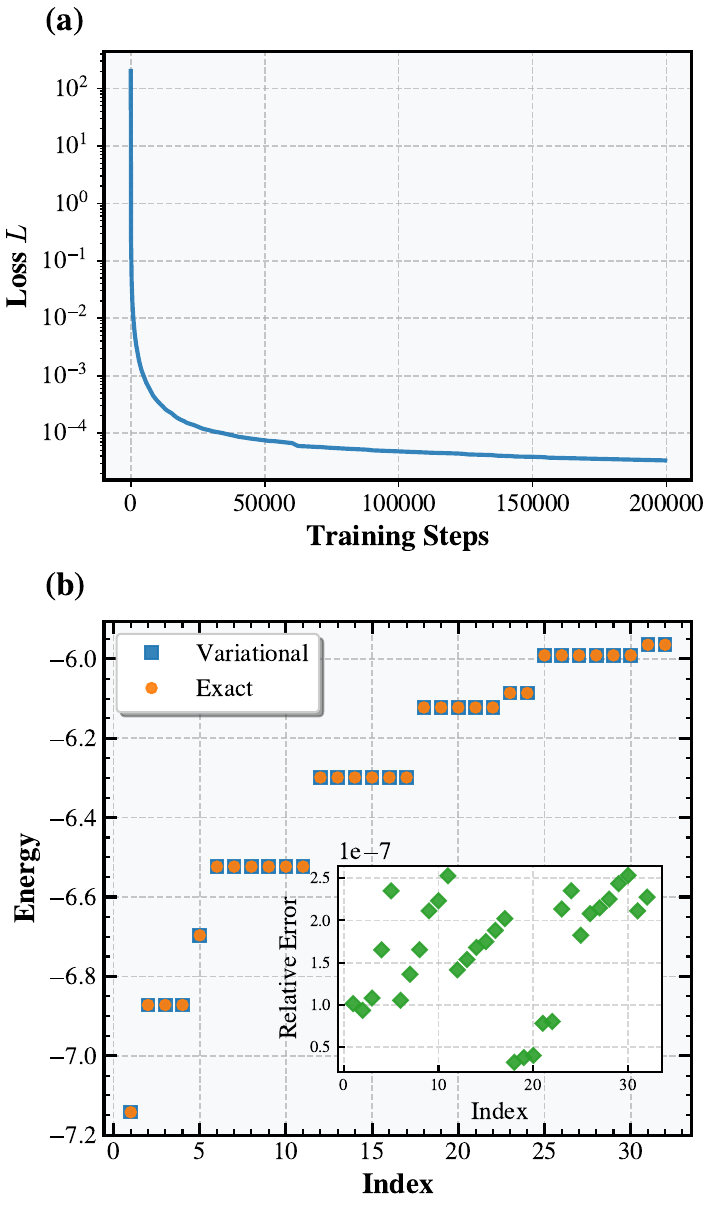}
	\caption{(a) Convergence of the loss function $L = \Tr(\mathbf{S}^{-1}\mathbf{H})$ with training steps for the low-lying excited state problem in 1D spin-half chain in Eq.~\eqref{eq:heis} with size $N=16$ and $J_x=J_y=J_z=1$, $h_z=0$. The loss is shifted by the sum of exact eigenvalues for the logarithmic scale. (b) Comparison of the low-energy spectrum obtained from our variational method (blue squares) and exact results (orange circles). The lowest $N_s=32$ energy levels are shown, calculated using $32$ variational MPS with bond dimension $\chi=16$. The inset shows the relative error $|E_{vari} - E_{exact}| / |E_{exact}|$ for each energy level (green diamonds), demonstrating high accuracy with relative errors on the order of $10^{-7}$. 
}
\label{fig:mps_main}
\end{figure}

Fig.~\ref{fig:mps_main}(a) shows the convergence of the loss function $L(\vec{\theta})$ during the optimization process. The loss function decreases steadily, indicating that the set of variational MPS is converging towards spanning the subspace of the lowest energy eigenstates. It is worth noting that the loss can be further optimized given more training budgets as the convergence is not fully reached. Fig.~\ref{fig:mps_main}(b) presents
the calculated energies $E_1, E_2, \dots, E_{32}$ obtained by solving the generalized eigenvalue problem $\mathbf{H}c = E \mathbf{S}c$ after the optimization. We compare these energies to benchmark results obtained from exact diagonalization,
and find the variational energies are in excellent agreement with the exact values, demonstrating that our method successfully captures the lowest $N_s$ energy levels of the system. The results here and additional results for different Hamiltonian parameters and more training budgets \cite{Note1} demonstrate that we are able to variationally obtain these states simultaneously using a set of non-orthogonal non-normalized MPS without further numerical tricks or hyperparameter tuning for numerical stability. It is also straightforward to generalize the approach to projected entangled-pair states~\cite{Verstraete2004, Verstraete2004a} where spectrum problem in two dimensions can be variationally optimized, though the complexity of exactly calculating overlap elements is not scalable.

\textit{Results for Morse oscillator spectrum with quantics tensor train.---} \label{sec:morse_qtt} Next, we apply our variational principle to find the vibrational excited states of a diatomic molecule modeled by a one-dimensional Morse potential $V(x) = D_e(1 - e^{-a_M(x-r_e)})^2$, where $x$ is the displacement from the equilibrium bond distance $r_e$. We use parameters representative of an O-H stretching vibration: dissociation energy $D_e = 42301 \text{ cm}^{-1}$, Morse parameter $a_M = 2.1440 \text{ \AA}^{-1}$, and reduced mass $\mu =0.9527 \text{ amu}$ \cite{Vogt2019, Vogt2025}. The Hamiltonian $H = -\frac{\hbar^2}{2\mu} \frac{d^2}{dx^2} + V(x)$ is defined on a discretized grid of $2^{N_d}$ points spanning the relevant range of the coordinate $x$, representing the discretized version of the 1D real space. For a total of $2^{N_d}$ grid points, a coordinate value $x$ on the range $[0, 10\text{ \AA}]$ is mapped to a unique $N_d$-bit index $x_1 x_2 \dots x_N \in \{0,1\}^{N_d}$, such that $x = \sum_{k=1}^N x_k 2^{-k} $ in the unit of $10\text{ \AA}$. For $N_d=16$, the mesh size is approximately $1.526\times 10^{-4}\text{ \AA}$ which gives a discretization relative error around $1.5\times 10^{-7}$ \cite{Note1}.

We represent the excited state wavefunction on this grid index space using the quantics tensor train format which is a matrix product state representation for discretizations of continuous variables. The wavefunction, viewed as a tensor $\Psi_{x_1 x_2 \dots x_{N_d}}$ of order $N_d$, is decomposed into a product of $N$ core tensors $\mathbf{A}_k(x_k)$:
$\Psi_{x_1 x_2 \dots x_{N_d}} = \mathbf{A}_1(x_1) \mathbf{A}_2(x_2) \dots \mathbf{A}_{N_d}(x_{N_d})$,
where $\mathbf{A}_k(x_k)$ are rank-3 tensors with indices connecting neighboring sites (bits), $\mathbf{A}_1(x_1)$ and $\mathbf{A}_N(x_{N_d})$ effectively being vectors. The dimensions of these connecting indices define the TT ranks $\chi$. Each variational state $|\psi_i\rangle$ is approximated by an independent TT representation parameterized by its core tensors $\{\mathbf{A}_{i,k}(x_k)\}$, with a specified maximal TT rank $\chi=128$. We target the lowest $N_s=16$ vibrational states of the potential.

Calculating the overlap matrix elements $\mathbf{S}_{ij} = \langle\psi_i | \psi_j\rangle$ and Hamiltonian matrix elements $\mathbf{H}_{ij} = \langle\psi_i | H | \psi_j\rangle$ for TT wavefunctions on a grid involves contracting the TT representations of the states and the operators. Both the potential energy operator and the kinetic energy operator can in general be modeled by a matrix product operator \cite{Jolly2023}. The required matrix elements can also be efficiently obtained via tensor network contraction with $O(N_dN_s^2\chi^3)$ complexity. 

\begin{figure}[htbp]
\centering
\includegraphics[width=0.92\columnwidth]{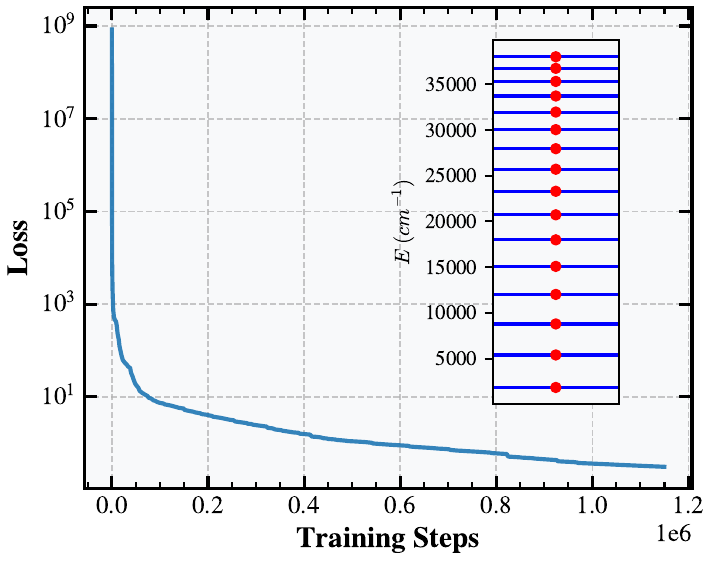} 
\caption{Training dynamics for the Morse potential spectrum application ($N_s=16$, $N_d=16$, $\chi=128$). The loss is shifted by the exact sum of $N_s$ lowest eigenenergies for the logarithmic scale. The inset shows the variational eigenenergies (red point) against the exact value (blue line) which are in agreement.} 
\label{fig:morse_convergence}
\end{figure}

Fig.~\ref{fig:morse_convergence} shows the typical convergence behavior of the loss function during the optimization.
The variational energies are in excellent agreement with the analytical values with relative errors of the order of $10^{-5}$ \cite{Note1}, showcasing our method's accuracy for this continuous-variable problem. 
This application demonstrates the flexibility of our method in handling wavefunctions defined on continuous coordinates. Combined with the compressive power of quantics tensor train formats, the method provides an efficient way to obtain multiple vibrational excited states simultaneously, which is valuable for spectroscopic calculations.

\begin{figure}[htbp]
\centering
\includegraphics[width=0.9\columnwidth]{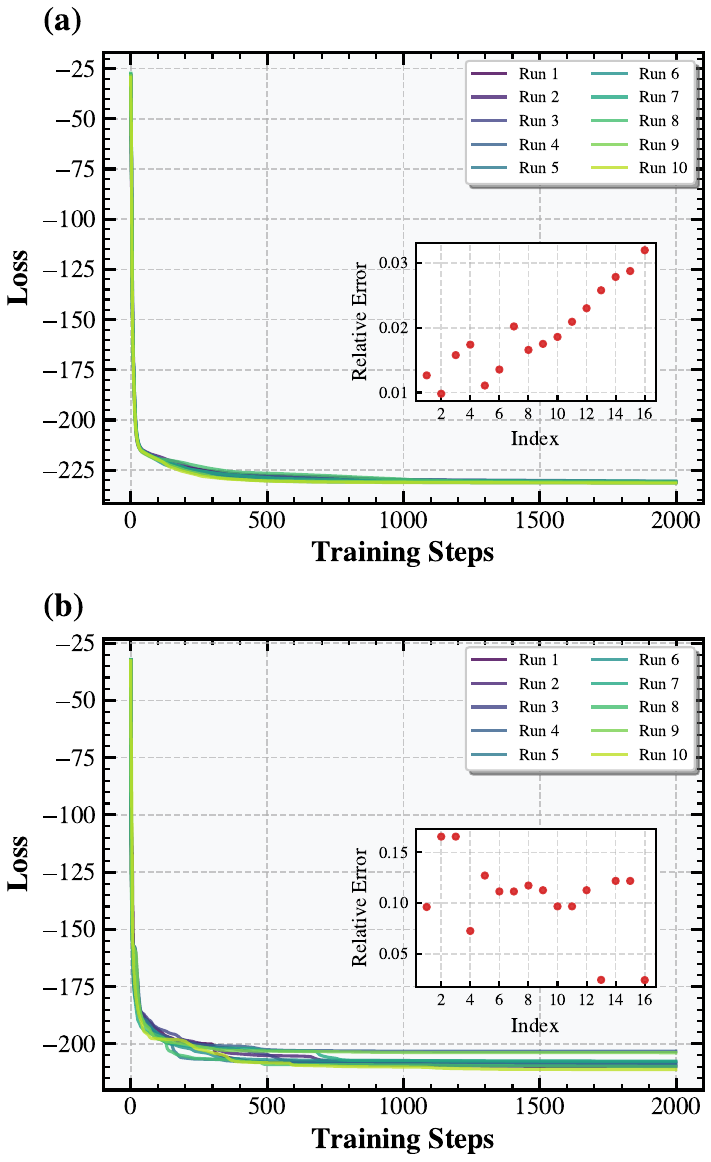} 
\caption{Training dynamics for the VQE calculation of the 2D Hubbard model ($N=12$ qubits, $N_s=16$ states, $D=5$ hardware effcient ansatz). The inset plot shows the relative energy error obtained from the best of 10 optimization trials. (a) Our method. (b) Subspace VQE.}
\label{fig:hubbard_convergence}
\end{figure}

\textit{Results for 2D fermionic Hubbard model with variational quantum circuits.---} \label{sec:vqe}
Finally, we explore the application of our method to strongly correlated fermionic systems, specifically the two-dimensional Hubbard model on a $2 \times 3$ square lattice with open boundary conditions. The Hubbard Hamiltonian is given by
\begin{equation}
H = -t \sum_{\langle i, j \rangle, \sigma} (c_{i\sigma}^\dagger c_{j\sigma} + c_{j\sigma}^\dagger c_{i\sigma}) + U \sum_i (n_{i\uparrow}-\frac{1}{2}) (n_{i\downarrow}-\frac{1}{2}),
\end{equation}
where $c_{i\sigma}^\dagger$ ($c_{i\sigma}$) creates (annihilates) an electron at site $i$ with spin $\sigma \in \{\uparrow, \downarrow\}$, $n_{i\sigma} = c_{i\sigma}^\dagger c_{i\sigma}$ is the number operator, $t$ is the hopping amplitude, and $U$ is the on-site repulsion. We use $t=1$ and $U=4$ throughout the calculation. The sum $\langle i, j \rangle$ runs over nearest-neighbor sites. Using Jordan-Wigner transformation, this Hamiltonian is mapped to a sum of Pauli strings acting on $N=12$ qubits. 

We employ a set of $N_s=16$ independent variational quantum circuits (VQCs) to prepare the variational states $\{|\psi_i(\vec{\theta}_i)\rangle = U(\vec{\theta}_i)|0\dots0\rangle\}$. Each VQC $U(\vec{\theta}_i)$ utilizes a hardware-efficient ansatz consisting of $D=5$ repeated blocks. Each block $d \in \{1, \dots, D\}$ consists of:
a layer of single-qubit rotations where $R_y,R_z, R_x$ rotations are applied and a layer of two-qubit parameterized entanglers applied on each adjacent pair of qubits $(j, j+1)$ including $e^{i\theta_{zz, d,j} Z_j Z_{j+1}}$, $e^{i\theta_{xx, d,j} X_j X_{j+1}}$, and $e^{i\theta_{yy, d,j} Y_j Y_{j+1}}$. 
The total number of variational parameters is of the order $O(NN_sD)$. The matrix elements $\mathbf{S}_{ij} $ and $\mathbf{H}_{ij} $ can be measured by Hadamard test or more efficient protocols on quantum computers \cite{Huggins2020}.

We benchmark the performance of our variational method against subspace variational quantum eigensolver (VQE) approach for excited states \cite{Nakanishi2019}. In subspace VQE, distinct input product states $\{|s_k\rangle\}$ are fed into the same circuit $V(\vec{\phi})$ to produce a set of variational states $\{|\tilde{\psi}_k(\vec{\phi})\rangle = V(\vec{\phi})|s_k\rangle\}$ that are encouraged to approximate the low-energy eigenstates and are mutually orthogonal by default. The optimization for subspace VQE minimizes a loss function including the sum of energy expectation values. We used the same ansatz for the circuit $V(\vec{\phi})$ as in our method's individual PQCs and targeted the lowest $N_s$ states for comparisons. 
For each method, we conducted 10 independent optimization trials with different random initial parameters. Each trial ran for 2000 optimization steps using the L-BFGS optimizer. We compare the results of the best-performing trial for each method.

Fig.~\ref{fig:hubbard_convergence}(a) shows the convergence of the loss function $L$ for our method using the VQE ansatz. The converged loss in our method is much smaller than subspace VQE in Fig.~\ref{fig:hubbard_convergence}(b) and the variance of the loss across different trials is also smaller in our case, indicating the more robust training landscape in our variational method. The resulting excited state energy estimation shown in the insets also supports the advantage of our method.
This application showcases the potential of our variational principle on quantum computers.

{\textit{Discussions.---}} 
We have introduced and demonstrated a novel variational principle for simultaneously calculating multiple low-energy excited states of quantum many-body systems. Despite its promise, the method also presents challenges that warrant further investigation. The computation of matrix elements $\langle\psi_i|H|\psi_j\rangle$ and $\langle\psi_i|\psi_j\rangle$ and their gradients remains the primary computational bottleneck, heavily dependent on the chosen ansatz. This could be demanding for highly complex ansatzes or on quantum hardware with limited measurement resources. Furthermore, the optimization landscape of $L$ is non-convex, and convergence to the global minimum is not guaranteed, as is typical for variational methods. Numerical stability related to the overlap matrix $\mathbf{S}$ needs careful monitoring, although in our numerical experiments, the condition number of $\mathbf{S}$ remains manageable, and standard optimization techniques are surprisingly robust without explicit regularization. These facts suggest that the optimization sufficiently encourages the states to span a diverse enough subspace instead of collapsing towards the same low-energy subspace with
 near-singular $\mathbf{S}$ \cite{Note1}. 

We note the potential for integrating the variational principle with neural quantum states (NQS)~\cite{Carleo2017}, particularly within the variational Monte Carlo (VMC) framework. While the 
 loss function $L$ could in principle be directly applied with NQS ansatz, a more efficient, numerically robust and mathematically equivalent formulation has recently been proposed and successfully utilized in VMC calculations with NQS \cite{Pfau2024}. 
Adapting the evaluation of the loss function to leverage the unique structure of given ansatzes represents a promising avenue for future research to maximize the practical benefits of the proposed variational framework.

\textit{Acknowledgement.---} We acknowledge helpful discussions with Zi-Xiang Li, Shang Liu, Wei Tang, and Yantao Wu. LW is supported by the National Natural Science Foundation of China under Grants No. T2225018, No. 92270107, No. T2121001, and National Key Projects for Research and Development of China Grant. No. 2021YFA1400400. SXZ is supported by a start-up grant at IOP-CAS.

\let\oldaddcontentsline\addcontentsline
\renewcommand{\addcontentsline}[3]{}
%

\let\addcontentsline\oldaddcontentsline
\onecolumngrid

\clearpage
\newpage
\widetext

\begin{center}
\textbf{\large Supplemental Material for ``A Unified Variational Framework for Quantum Excited States''}
\end{center}

\date{\today}
\maketitle

\renewcommand{\thefigure}{S\arabic{figure}}
\setcounter{figure}{0}
\renewcommand{\theequation}{S\arabic{equation}}
\setcounter{equation}{0}
\renewcommand{\thesection}{\Roman{section}}
\setcounter{section}{0}
\setcounter{secnumdepth}{4}

\addtocontents{toc}{\protect\setcounter{tocdepth}{0}}
{
\tableofcontents
}

\section{Rayleigh-Ritz principle for excited states}

This section provides a more formal justification for the variational nature of the energies obtained by solving the generalized eigenvalue problem $\mathbf{H}c = E \mathbf{S}c$ within a finite-dimensional subspace, specifically demonstrating that these energies serve as upper bounds to the exact eigenvalues of the Hamiltonian. The theoretical foundation for this is the minimax principle, also known as the Rayleigh-Ritz theorem.

Let $H$ be a self-adjoint operator (Hamiltonian) acting on a Hilbert space $\mathcal{H}$. Its eigenvalues are $E_1^{exact} \le E_2^{exact} \le \dots$, with corresponding orthonormal eigenstates $|\Psi_1^{exact}\rangle, |\Psi_2^{exact}\rangle, \dots$.

For any non-zero state $|\psi\rangle \in \mathcal{H}$, the Rayleigh quotient is defined as:
$$ R(|\psi\rangle) = \frac{\langle\psi | H | \psi\rangle}{\langle\psi | \psi\rangle}. $$
The basic variational principle states that $R(|\psi\rangle) \ge E_1^{exact}$ for any $|\psi\rangle$, and the minimum is achieved when $|\psi\rangle = |\Psi_1^{exact}\rangle$.

The minimax principle extends this to all eigenvalues. It characterizes the $k$-th eigenvalue $E_k^{exact}$ (starting from $k=1$) in two equivalent ways:

\begin{align*} E_k^{exact} &= \min_{\substack{W \subseteq \mathcal{H} \\ \dim(W) = k}} \max_{\substack{|\psi\rangle \in W \\ |\psi\rangle \ne 0}} R(|\psi\rangle) \quad \text{(Min of Max)} \\ E_k^{exact} &= \max_{\substack{W \subseteq \mathcal{H} \\ \dim(W) = k-1}} \min_{\substack{|\psi\rangle \in W^\perp \\ |\psi\rangle \ne 0}} R(|\psi\rangle) \quad \text{(Max of Min)} \end{align*}
Here, $W$ is a subspace of $\mathcal{H}$, $\dim(W)$ is its dimension, and $W^\perp$ is its orthogonal complement. The first min of max formulation is particularly useful here. It shows that if you consider all possible $(k)$-dimensional subspaces $W$ of the full Hilbert space, find the maximum value of the Rayleigh quotient within each subspace, the minimum among these maximum values gives the exact $k$-th eigenvalue $E_k^{exact}$.

Now consider our variational method. We choose a set of $N_s$ linearly independent variational states $\{|\psi_i\rangle\}_{i=1}^{N_s}$. These states span a specific $N_s$-dimensional subspace $V = \text{span}\{|\psi_1\rangle, \dots, |\psi_{N_s}\rangle\}$. By solving the generalized eigenvalue problem $\mathbf{H}c = E \mathbf{S}c$, we obtain $N_s$ eigenvalues $E_1 \le E_2 \le \dots \le E_{N_s}$. These are the Ritz values for the subspace $V$. The corresponding eigenvectors $c_\alpha$ define states $|\Psi_\alpha\rangle = \sum_i c_{i\alpha} |\psi_i\rangle$ that form an orthonormal basis for $V$ and diagonalize $H$ within $V$, meaning $\langle\Psi_\alpha | H | \Psi_\beta\rangle = E_\alpha \delta_{\alpha\beta}$ for $|\Psi_\alpha\rangle, |\Psi_\beta\rangle \in V$.

The Ritz values $E_k$ obtained from the subspace $V$ can also be characterized by a minimax principle, but restricted to subspaces within $V$:
$$ E_k = \min_{\substack{W' \subseteq V \\ \dim(W') = k}} \max_{\substack{|\psi\rangle \in W' \\ |\psi\rangle \ne 0}} R(|\psi\rangle), \quad \text{for } k = 1, 2,\dots, N_s. $$
This states that the $k$-th Ritz value $E_k$ is the minimum among the maximum Rayleigh quotients over all $k$-dimensional subspaces $W'$ that are contained within the chosen $N_s$-dimensional variational subspace $V$.

Now we can prove that $E_k \ge E_k^{exact}$ for $k = 1, \dots, N_s$.
Compare the two minimax formulations:
\begin{align*} E_k^{exact} &= \min_{\substack{W \subseteq \mathcal{H} \\ \dim(W) = k}} \left( \max_{\substack{|\psi\rangle \in W \\ |\psi\rangle \ne 0}} R(|\psi\rangle) \right) \\ E_k &= \min_{\substack{W' \subseteq V \\ \dim(W') = k}} \left( \max_{\substack{|\psi\rangle \in W' \\ |\psi\rangle \ne 0}} R(|\psi\rangle) \right) \end{align*}
The set of subspaces $\{W' \subseteq V \mid \dim(W') = k\}$ is a subset of the set of all subspaces $\{W \subseteq \mathcal{H} \mid \dim(W) = k\}$. Since we are taking the minimum value over a set, the minimum over a restricted set (subspaces $W'$ within $V$) must be greater than or equal to the minimum over the larger set (all subspaces $W$ in $\mathcal{H}$).
Therefore, for all $k = 1, \dots, N_s$:
$$ E_k \ge E_k^{exact}. $$
This proves that each individual Ritz value $E_k$ obtained from the generalized eigenvalue problem within the variational subspace $V$ provides an upper bound to the corresponding exact eigenvalue $E_k^{exact}$.

Our variational method minimizes the sum $\sum_{\alpha=1}^{N_s} E_\alpha(\vec{\theta})$. By minimizing this sum of upper bounds with respect to the variational parameters $\vec{\theta}$, the method effectively seeks to find the subspace $V_{\vec{\theta}}$ whose Ritz values are as close as possible to the true lowest $N_s$ exact eigenvalues, while always satisfying the upper bound property $E_\alpha(\vec{\theta}) \ge E_\alpha^{exact}$ for each eigenvalue.

\section{Generalized eigenvalue problems}

Since $\mathbf{S}$ is positive definite and thus invertible, the loss function form is equivalent to the sum of eigenvalues obtained from solving $\mathbf{S}^{-1}\mathbf{H}c = Ec$. The eigenvalues of the matrix $\mathbf{S}^{-1}\mathbf{H}$ are precisely the generalized eigenvalues $E_\alpha$ of the subspace problem and the trace of the matrix is exactly the sum of its eigenvalues. Therefore, $L = \Tr(\mathbf{S}^{-1}\mathbf{H}) = \sum_{\alpha=1}^{N_s} E_\alpha$, where $E_\alpha$ are the approximate energies obtained from the current set of variational states $\{|\psi_i\rangle\}$. By minimizing this sum, the optimization process drives the variational states $\{|\psi_i\rangle\}$ to span the subspace corresponding to the lowest possible sum of $N_s$ energy levels, which will approximate the lowest $N_s$ energy levels of the full Hamiltonian if the ansatz is sufficiently expressive.

This standard generalized eigenvalue problem can transformed into a standard symmetric eigenvalue problem at the post-processing stage. Since $\mathbf{S}$ is Hermitian and positive definite, it can be decomposed using the Cholesky decomposition:
\begin{equation}
\mathbf{S} = L L^\dagger,
\end{equation}
where $L$ is a lower triangular matrix and $L^\dagger$ is its conjugate transpose. Substituting this into the generalized eigenvalue equation:
\begin{equation}
\mathbf{H}c = E (L L^\dagger) c.
\end{equation}
We can make a substitution $v = L^\dagger c$. Since $L^\dagger$ is invertible, $c = (L^\dagger)^{-1} v$. Substituting this into the equation and multiplying by $L^{-1}$ from the left yields:
\begin{align}
L^{-1} \mathbf{H} (L^\dagger)^{-1} v &= E L^{-1} L L^\dagger (L^\dagger)^{-1} v \\
(L^{-1} \mathbf{H} (L^\dagger)^{-1}) v &= E v.
\end{align}
This is a standard symmetric eigenvalue problem $Mv = Ev$, where $M = L^{-1} \mathbf{H} (L^\dagger)^{-1}$. Since $\mathbf{H}$ is Hermitian, $M$ is also Hermitian.
Solving the standard eigenvalue problem for $M$ provides the eigenvalues $E_\alpha$ (which are the desired low-energy energies) and the eigenvectors $v_\alpha$. The original coefficient vectors $c_\alpha$ are then recovered by solving the triangular system $L^\dagger c_\alpha = v_\alpha$ for each $\alpha$.

\section{Relation to existing subspace optimization methods}\label{sec:relation_subspace}
Our method minimizes the sum of the generalized eigenvalues $\Tr(\mathbf{S}^{-1}\mathbf{H}) = \sum_\alpha E_\alpha$ within a subspace, which is a concept shared with certain existing techniques. However, our approach presents significant differences in its formulation and implementation, leading to distinct advantages.

Traditional subspace optimization methods, such as state-averaged multi-configurational self-consistent field (SA-MCSCF) \cite{Hinze1973, Diffenderfer1982}, also aim to minimize a weighted sum of energies (Ritz values) obtained within a defined subspace. In SA-MCSCF, this subspace is spanned by a set of configuration state functions (CSFs), which are fixed linear combinations of Slater determinants constructed from a set of molecular orbitals (MOs). The optimization procedure then iteratively refines the MOs (which non-linearly affect the CSF basis) and the linear coefficients of the states within the CSF basis (the CI coefficients). A key step in each optimization iteration involves solving a standard eigenvalue problem (CI step) within the current basis (the dimension of the matrix can be much larger than $N_s$ as in our case) to obtain the Ritz values and their corresponding wavefunctions. This iterative diagonalization within the optimization loop is characteristic of such methods.

Another class of related quantum algorithms focuses on constructing a subspace using a fixed or heuristically generated set of quantum states, and then solving the generalized eigenvalue problem within this static subspace. These ``quantum space expansion'' methods typically generate basis states by applying the related operators or time evolution to a reference state \cite{McClean2017, Parrish2019, Motta2020, Baek2023}. The quality of the resulting approximation to the low-energy spectrum depends entirely on the initial reference state and the chosen expansion operators. Once the basis is fixed, the matrix elements $\mathbf{H}$ and $\mathbf{S}$ are measured, and the generalized eigenvalue problem is solved once to obtain the approximate energies and states within that particular subspace. These methods do not typically involve a variational optimization loop to improve the subspace itself.

More recently, the concept of using a subspace spanned by a collection of non-orthogonal parameterized quantum states to solve a generalized eigenvalue problem has been explored in the context of variational quantum algorithms. Specifically, Ref.~\cite{Huggins2020} introduced a non-orthogonal variational quantum eigensolver (NOVQE) which utilizes such a subspace, spanned by independently parameterized quantum circuits $\{|\phi_i(\vec{\theta}_i)\rangle\}$. Like our approach, this involves measuring off-diagonal matrix elements $\langle\phi_i | H | \phi_j\rangle$ and $\langle\phi_i | \phi_j\rangle$ to form $\mathbf{H}$ and $\mathbf{S}$ matrices, and solving a generalized eigenvalue problem $\mathbf{H}c = E\mathbf{S}c$ within the subspace. However, a crucial distinction lies in the variational principle applied. The NOVQE method primarily focuses on minimizing the lowest eigenvalue to approximate the ground state energy from this generalized eigenvalue problem. A similar philosophy underlies a recent work \cite{Giuliani2025}, which applies this generalized eigenvalue problem based variational approach in quantum chemistry ground state problem using wavefunctions composed of multiple Slater determinants.

In contrast, our method introduces a distinct variational principle based on minimizing the trace $L = \Tr(\mathbf{S}^{-1}\mathbf{H})$. Our approach is designed to simultaneously optimize the variational states $\{|\psi_i\rangle\}$ to span the low-energy manifold corresponding to the lowest $N_s$ energy levels. This key difference in the objective function leads to the capability of finding multiple low-lying excited states concurrently, rather than primarily targeting the ground state. Furthermore, our method formulates the problem as a direct gradient-based optimization of the loss function $L(\vec{\theta})$, where the generalized eigenvalue problem is solved only as a post-processing step after parameter optimization, simplifying the integration with standard automatic differentiation tools compared to methods that require eigenvalue decomposition within the iterative optimization loop. And the implementation of our approach is also very friendly to computation intensive hardware such as GPU.

These differences highlight the novelty of our approach. By framing the problem as a direct variational optimization of $\Tr(\mathbf{S}^{-1}\mathbf{H})$ over the parameters of general non-orthogonal ansatzes, we provide a flexible framework applicable to a wide variety of quantum systems and state representations, circumventing the need for penalty terms or complex iterative diagonalization procedures within the main optimization loop, and crucially, enabling the simultaneous discovery of multiple excited states.

It is worth noting that a very similar form of loss function is employed in Refs.~\cite{Ordejn1993, Yang1997}.
While our variational principle $\Tr(\mathbf{S}^{-1}\mathbf{H})$ bears a superficial resemblance to the loss functions in Refs. \cite{Ordejn1993, Yang1997} in the context of linear-scaling electronic structure methods, the fundamental goals, the nature of the variational objects, and the role of the inverse overlap are distinctly different. Previous works focus on ground state electronic structure calculations by representing the one-electron density matrix using a number of localized one-electron orbitals ($M$) whose number scales with the system size. The key challenge for linear scaling in such methods is avoiding the explicit computation of the $M \times M$ overlap matrix $\mathbf{S}$ inverse, which scales cubically with system size. In stark contrast, our method targets the simultaneous determination of a fixed, small number ($N_s$) of many-body excited states for general quantum systems. Our variational objects are the full many-body wavefunctions $\{|\psi_i\rangle\}_{i=1}^{N_s}$. The overlap matrix $\mathbf{S}$ is an $N_s \times N_s$ small matrix of overlaps between these many-body states. Crucially, the size $N_s$ is determined by the desired number of excited states and is entirely independent of the system size. Therefore, the explicit calculation of the small $N_s \times N_s$ matrix inverse $\mathbf{S}^{-1}$ at each optimization step is computationally trivial and does not introduce any scaling impediment for large systems. Our principle is tailored for variational excited state calculations without explicit orthogonality constraints, which is a distinct problem from ground state methods based on one-electron density matrices as in Refs.~\cite{Ordejn1993, Yang1997}.

\section{More numerical results for 1D spin chain with MPS}

This section provides additional numerical results for the 1D spin chain application with MPS ansatz discussed in the main text.

\begin{figure}[t]\centering
	\includegraphics[width=0.48\textwidth]{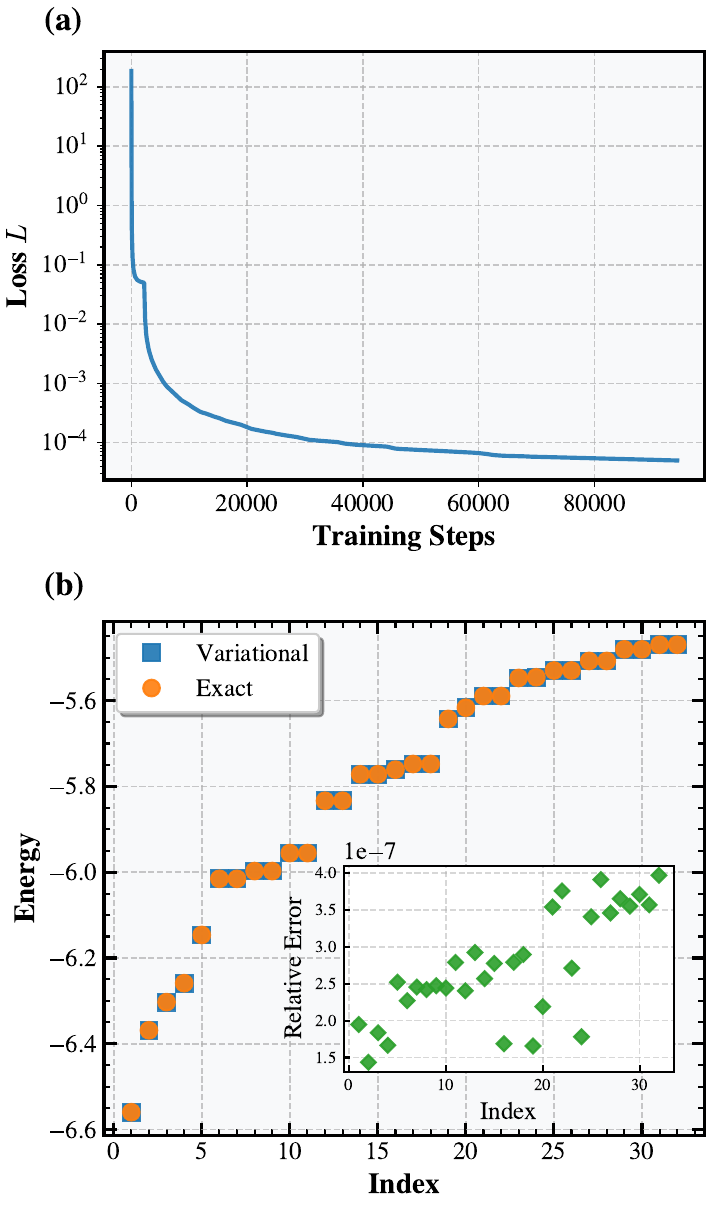}
	\caption{(a) Convergence of the loss function $L = \Tr(\mathbf{S}^{-1}\mathbf{H})$ with training steps for the low-lying excited state problem in 1D spin chain in Eq.~\eqref{eq:heis} with size $N=16$ and $J_x=1, J_y=0.95, J_z=0.8$, $h_z=0.015$. (b) Comparison of the low-energy spectrum obtained from our variational method (blue squares) and exact results (orange circles). The lowest 32 energy levels are shown, calculated using $N_s=32$ variational MPS with bond dimension $\chi=16$. The inset shows the relative error $|E_{vari} - E_{exact}| / |E_{exact}|$ for each energy level (green diamonds), demonstrating high accuracy with relative errors on the order of $10^{-7}$. The close agreement between variational and exact energies across multiple states confirms the method's ability to simultaneously find low-lying excited states.}
\label{fig:heisenberg_aniso_results}
\end{figure}

\begin{figure}[t]\centering
	\includegraphics[width=0.48\textwidth]{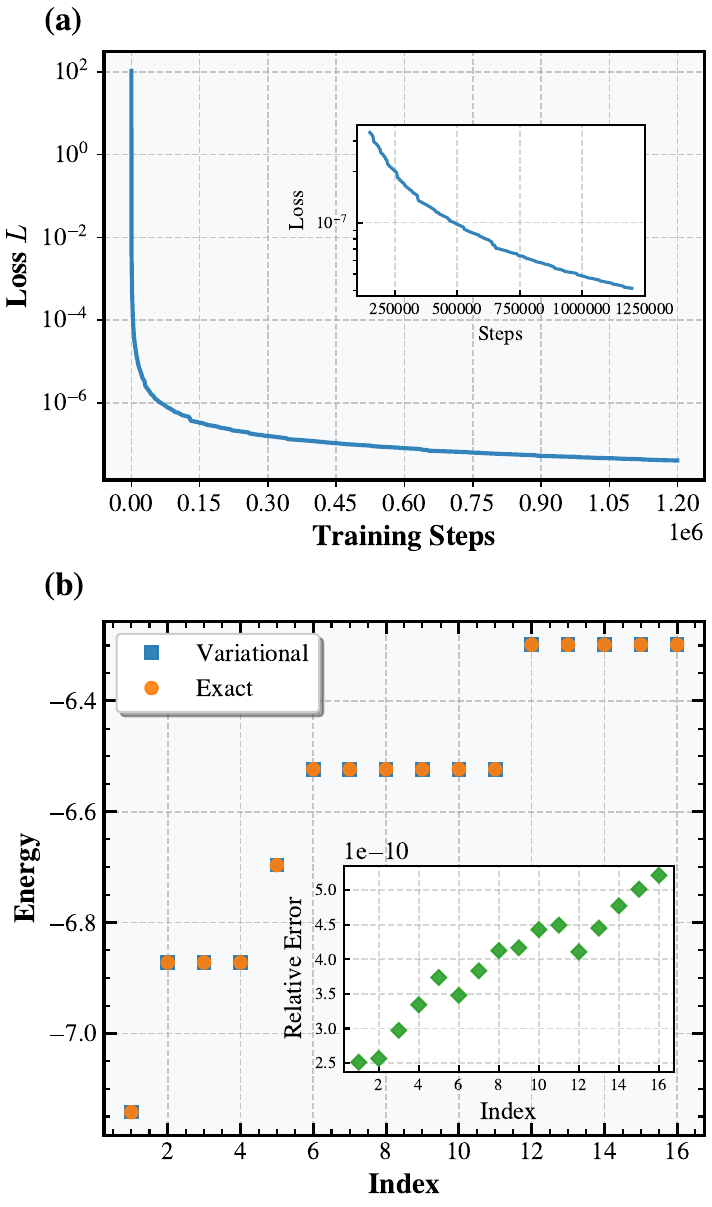}
	\caption{(a) Convergence of the loss function $L = \Tr(\mathbf{S}^{-1}\mathbf{H})$ with training steps for the low-lying excited state problem in 1D spin chain in Eq.~\eqref{eq:heis} with size $N=16$ and $J_x=J_y=J_z=1$, $h_z=0$. The inset shows the zoom-in training dynamics of the loss from $150000$ step. It is obvious that the accuracy can further get improved with additional training budgets. (b) Comparison of the low-energy spectrum obtained from our variational method (blue squares) and exact results (orange circles). The lowest 32 energy levels are shown, calculated using $N_s=16$ variational MPS with bond dimension $\chi=24$. The inset shows the relative error $|E_{vari} - E_{exact}| / |E_{exact}|$ for each energy level (green diamonds), demonstrating high accuracy with relative errors on the order of $10^{-10}$ which matches the result reported in Ref. \cite{Pirvu2012}.}
\label{fig:heisenberg_iso_results_chi24}
\end{figure}

Fig.~\ref{fig:heisenberg_aniso_results} and \ref{fig:heisenberg_iso_results_chi24} show the training curves and accuracy of the optimized energy spectra for the anisotropic ($\chi=16$) and isotropic ($\chi=24$) Heisenberg models, respectively. The former result demonstrates that our approach is applicable for different Hamiltonian parameters. In terms of the case, we use $N_s=16$ variational states, slightly larger bond dimension $\chi=24$, and many more training steps. The loss is still decreasing at the late training stage as shown in \ref{fig:heisenberg_iso_results_chi24}(a), indicating potential for even better accuracy with additional training budgets reported here. And the relative energy error $O(10^{-10})$ matches the state-of-the-art results reported in Ref.~\cite{Pirvu2012}.

We find that the relative error follows a linear scaling with respect to the eigenvalue index after sufficient variational training. By fitting, we find the relative error for $k$-th eigenvalue $\text{Relative Error}_k \approx 1.64\times 10^{-11}\times k + 2.56\times 10^{-10}$. By extrapolation, the relative error for $N_s=160$ eigenstates is still on the order of $O(10^{-9})$, suggesting that the method can accurately capture a large number of eigenstates with reasonable bond dimensions.

\begin{figure}[htbp]
\centering
\includegraphics[width=0.45\columnwidth]{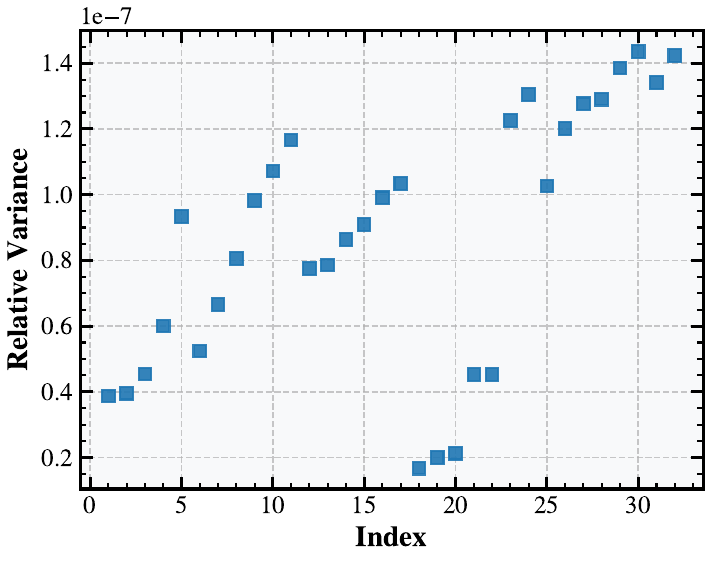} 
\caption{Relative energy variance of the calculated states for the 1D Heisenberg chain application with size $N=16$, Hamiltonian parameter $J_x=J_y=J_z=1$, $h_z=0$ and MPS bond dimension $\chi=16$. The plot shows the relative variance for the lowest $N_s=32$ calculated states, indexed from $k=1$ to $32$. The relative variance for all states is on the order of $10^{-7}$, which is very small, indicating that the obtained states are high-quality approximations to the eigenstates of the Hamiltonian.}
\label{fig:iso_heisenberg_variance_sm} 
\end{figure}

\begin{figure}[htbp]
\centering
\includegraphics[width=0.45\columnwidth]{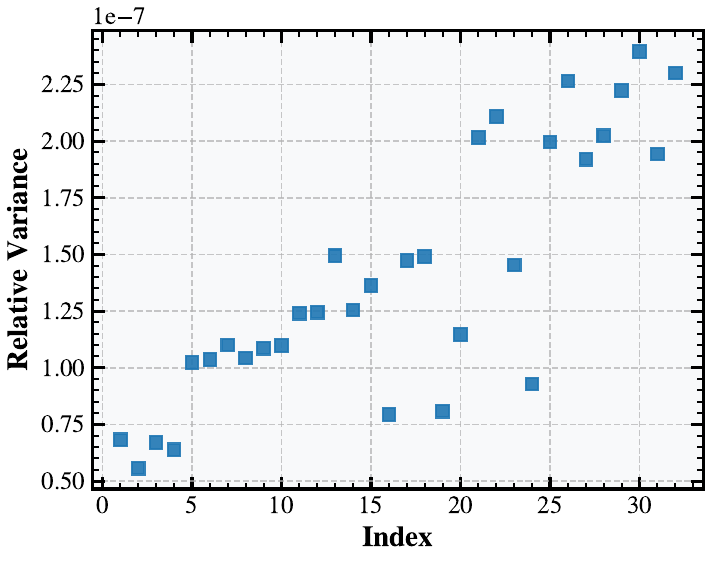} 
\caption{Relative energy variance of the calculated states for the 1D Heisenberg chain application with size $N=16$, Hamiltonian parameter $J_x=1.0$, $J_y=0.95$, $J_z=0.8$, $h_z=0.015$ and MPS bond dimension $\chi=16$. 
The plot shows the relative variance for the lowest $N_s=32$ calculated states, indexed from $k=1$ to $32$. The relative variance for all states is on the order of $10^{-7}$, which is very small, indicating that the obtained states are high-quality approximations to the eigenstates of the Hamiltonian.}
\label{fig:heisenberg_variance_sm}
\end{figure}

\begin{figure}[htbp]
\centering
\includegraphics[width=0.45\columnwidth]{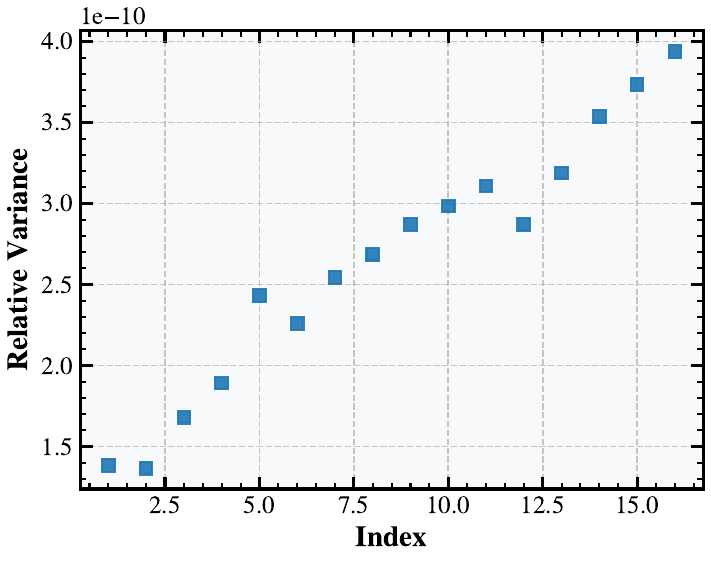} 
\caption{Relative energy variance of the calculated states for the 1D Heisenberg chain application with size $N=16$, Hamiltonian parameter $J_x=J_y=J_z=1$, $h_z=0$, and MPS bond dimension $\chi=24$. The plot shows the relative variance for the lowest $N_s=16$ calculated states, indexed from $k=1$ to $16$. The relative variance for all states is on the order of $10^{-10}$, indicating that the obtained states are high-quality approximations to the eigenstates of the Hamiltonian. The linear scaling between relative energy variance and eigenstate index is clearly shown in the plot.}
\label{fig:heisenberg_variance_chi24_sm} 
\end{figure}

For a state $|\Psi_k\rangle$ with calculated energy $E_k = \langle\Psi_k|H|\Psi_k\rangle$, the energy variance is defined as $\text{Var}(H)_k = \langle\Psi_k|H^2|\Psi_k\rangle - \langle\Psi_k|H|\Psi_k\rangle^2$. For an exact eigenstate, this variance is exactly zero. A small variance indicates that the obtained state is a high-quality approximation to an exact eigenstate. Importantly, this probe is scalable to large size systems where the exact eigenenergies cannot be reliably obtained for benchmark. We report the relative variance, defined as $\frac{\text{Var}(H)_k}{\langle\Psi_k|H|\Psi_k\rangle^2}$ for different Hamiltonian and ansatz parameters in Fig.~\ref{fig:iso_heisenberg_variance_sm}, Fig.~\ref{fig:heisenberg_variance_sm} and Fig.~\ref{fig:heisenberg_variance_chi24_sm}. In all cases, the relative variance is very small, indicating that the obtained states are excellent approximations to exact eigenstates of the Hamiltonian. For $\chi=16$, the relative variance is on the order of $10^{-7}$. Increasing the bond dimension to $\chi=24$ and with more training budgets further reduces the variance to $10^{-10}$, demonstrating the ability to achieve very high fidelity eigenstates by increasing the ansatz capacity. Interestingly, we find the relative variance also roughly follows a linear scaling with respect to the eigenstate index, this scaling is notably clearer for the better trained case as shown in Fig.~\ref{fig:heisenberg_variance_chi24_sm}.

\section{More numerical results for Morse potential spectrum}

For the Hamiltonian $H = -\frac{\hbar^2}{2\mu} \frac{d^2}{dx^2} + V(x)$ on a discretized grid of $2^{N_d}$ points, the wavefunction is a vector of size $2^{N_d}$. In the quantics TT representation, this vector is represented as a tensor of order $N_d$ with local indices corresponding to the bits of the grid point index. Explicitly constructing matrix product operators (MPOs) for the derivative and potential operators and contracting them with the TT representation would be the standard approach for large $N_d$. However, for the moderate number of bits $N_d=16, 18$ used in our results, the full vector size $2^{N_d} $ is still manageable. Therefore, for simplicity in implementation, we did not explicitly construct derivative and potential MPOs. Instead, we first contracted the TT representation of the variational state to obtain the full wavefunction vector on the grid. We then directly applied the discrete derivative operator (using finite differences) and the potential operator (multiplication by the potential value at each grid point) to this vector to compute the required expectation values and matrix elements $\langle\psi_i | H | \psi_j\rangle$. 

We investigated the impact of grid resolution by performing optimizations with $N_d=18$ bits, which corresponds to a finer discretization. However, we observed that the accuracy for the spectrum did not systematically improve, and the training sometimes became less stable with this finer discretization. The mechanism behind this behavior warrants further investigation.

Note that the error induced by the truncation of the real space $[0, 10 \text{\AA}]$ and the discretization of finite grid size is relatively small compared to the variational error of the TT ansatz. We quantified this discretization error by comparing the energies obtained from numerical diagonalization of the discretized Schr\"odinger equation on grids of different sizes to the standard analytical formula for Morse potential energies, $E_n = \omega (n+\frac{1}{2}) -\frac{(w(n+\frac{1}{2}))^2}{4D} $, where $\omega=a_M \sqrt{2D_e}$. For the ground state ($n=0$), the relative error induced by discretizing with $N_d=16$ bits is approximately $1.5\times10^{-7}$, and with $N_d=18$ bits it is $9.3\times 10^{-9}$. These discretization errors are significantly smaller than the relative variational errors obtained in our TT calculations (on the order of $10^{-5}$), confirming that the accuracy is limited by the expressiveness and optimization of the TT ansatz, not by the grid resolution.

\begin{figure}[t]\centering
	\includegraphics[width=0.5\textwidth]{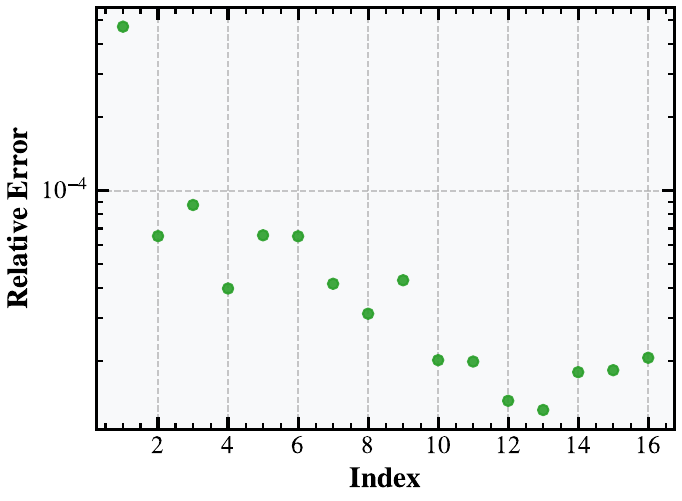}
	\caption{Relative energy error obtained from variational optimization on Morse potential with $N_d=16, N_s=16, \chi=128$ for the quantics TT ansatz.}
\label{fig:morse_error}
\end{figure}

The relative energy error for the variational solution in the main text is shown in Fig.~\ref{fig:morse_error} with order of $10^{-5}$. Again, the accuracy can be further improved with additional training steps as the training curve is still steadily decreasing after more than $10^6$ training steps.

\section{Hyperparameters and implementation details}

\textbf{Computational Setup:} All numerical simulations were performed using the \textsf{TensorCircuit-NG} library on an NVIDIA A800 80G GPU.

\textbf{Variational Ansatz:} We employed MPS ansatzes for the first two applications.
For the 1D spin chain, a periodic MPS was used.
For the Morse potential spectrum, an open-boundary MPS for quantics tensor train was used.
We note that for a fixed bond dimension $\chi$, periodic MPS generally has higher expressiveness and contraction time complexity than open-boundary MPS, while their space complexity is comparable.

\textbf{Optimization:}
\begin{itemize}
\item Optimizer choice: The L-BFGS optimizer was used for all applications, utilizing default settings provided by {\sf Optax} without hyperparameter tuning.
\item Stability: Training dynamics were generally stable, with no numerical issues encountered during the inversion of matrix $\mathbf{S}$. Condition numbers for the optimized $\mathbf{S}$ matrix were typically $O(10^2)$ (1D spin chain), $O(10^4)$ (Morse potential), and $O(10^2)$ (2D Hubbard model). The higher condition number in the Morse potential case correlated with occasional training instability observed in the experiments.
\item Convergence: The late-stage training convergence is in general notably slow. Preconditioners based on the quantum Fisher information matrix did not provide significant acceleration based on our experiments.
\item Alternative optimizers: First-order methods like Adam were found to be highly sensitive to learning rate schedules, frequently leading to optimization failure, particularly for the MPS-based applications.
\end{itemize}

\textbf{Parameter Initialization:} Initial variational parameters were randomly drawn from a Gaussian distribution with zero mean. The standard deviation ($\sigma$) specifically for each application was: $\sigma=1.0$ (1D spin chain), $\sigma=0.02$ (Morse potential), and $\sigma=0.01$ (2D Hubbard model).

\textbf{Performance:} Leveraging GPU acceleration, our approach demonstrated high efficiency. Typical execution times per optimization step on A800 GPU were around:
\begin{itemize}
\item $0.13s$ for the 1D spin chain application ($N=16, N_s=16, \chi=24$).
\item $0.09s$ for the Morse potential spectrum application ($N_d=18, N_s=16, \chi=256$).
\end{itemize}


\begin{thebibliography}{76}%
\makeatletter
\providecommand \@ifxundefined [1]{%
 \@ifx{#1\undefined}
}%
\providecommand \@ifnum [1]{%
 \ifnum #1\expandafter \@firstoftwo
 \else \expandafter \@secondoftwo
 \fi
}%
\providecommand \@ifx [1]{%
 \ifx #1\expandafter \@firstoftwo
 \else \expandafter \@secondoftwo
 \fi
}%
\providecommand \natexlab [1]{#1}%
\providecommand \bibnamefont  [1]{#1}%
\providecommand \bibfnamefont [1]{#1}%
\providecommand \citenamefont [1]{#1}%
\providecommand \href@noop [0]{\@secondoftwo}%
\providecommand \href [0]{\begingroup \@sanitize@url \@href}%
\providecommand \@href[1]{\@@startlink{#1}\@@href}%
\providecommand \@@href[1]{\endgroup#1\@@endlink}%
\providecommand \@sanitize@url [0]{\catcode `\\12\catcode `\$12\catcode `\&12\catcode `\#12\catcode `\^12\catcode `\_12\catcode `\%12\relax}%
\providecommand \@@startlink[1]{}%
\providecommand \@@endlink[0]{}%
\providecommand \url  [0]{\begingroup\@sanitize@url \@url }%
\providecommand \@url [1]{\endgroup\@href {#1}{\urlprefix }}%
\providecommand \urlprefix  [0]{URL }%
\providecommand \Eprint [0]{\href }%
\providecommand \doibase [0]{http://dx.doi.org/}%
\providecommand \selectlanguage [0]{\@gobble}%
\providecommand \bibinfo  [0]{\@secondoftwo}%
\providecommand \bibfield  [0]{\@secondoftwo}%
\providecommand \translation [1]{[#1]}%
\providecommand \BibitemOpen [0]{}%
\providecommand \bibitemStop [0]{}%
\providecommand \bibitemNoStop [0]{.\EOS\space}%
\providecommand \EOS [0]{\spacefactor3000\relax}%
\providecommand \BibitemShut  [1]{\csname bibitem#1\endcsname}%
\let\auto@bib@innerbib\@empty
\bibitem [{\citenamefont {Huber}\ and\ \citenamefont {Herzberg}(1979)}]{Huber1979}%
  \BibitemOpen
  \bibfield  {author} {\bibinfo {author} {\bibfnamefont {K.~P.}\ \bibnamefont {Huber}}\ and\ \bibinfo {author} {\bibfnamefont {G.}~\bibnamefont {Herzberg}},\ }Molecular spectra and molecular structure\ (\bibinfo  {publisher} {Springer US},\ \bibinfo {year} {1979})\BibitemShut {NoStop}%
\bibitem [{\citenamefont {Wölfle}(2018)}]{Wlfle2018}%
  \BibitemOpen
  \bibfield  {author} {\bibinfo {author} {\bibfnamefont {P.}~\bibnamefont {Wölfle}},\ }\bibfield  {title} {Quasiparticles in condensed matter systems,\ }\href {\doibase 10.1088/1361-6633/aa9bc4} {\bibfield  {journal} {\bibinfo  {journal} {Reports on Progress in Physics}\ }\textbf {\bibinfo {volume} {81}},\ \bibinfo {pages} {032501} (\bibinfo {year} {2018})}\BibitemShut {NoStop}%
\bibitem [{\citenamefont {Pal}\ and\ \citenamefont {Huse}(2010)}]{Pal2010a}%
  \BibitemOpen
  \bibfield  {author} {\bibinfo {author} {\bibfnamefont {A.}~\bibnamefont {Pal}}\ and\ \bibinfo {author} {\bibfnamefont {D.~A.}\ \bibnamefont {Huse}},\ }\bibfield  {title} {Many-body localization phase transition,\ }\href {\doibase 10.1103/PhysRevB.82.174411} {\bibfield  {journal} {\bibinfo  {journal} {Physical Review B - Condensed Matter and Materials Physics}\ }\textbf {\bibinfo {volume} {82}},\ \bibinfo {pages} {174411} (\bibinfo {year} {2010})},\ \bibinfo {note} {1010.1992The work give the rebirth of MBL by introducing spin half chain standard model in MBL and using ed to study the transition criticality and its fixed points}\BibitemShut {NoStop}%
\bibitem [{\citenamefont {Chen}\ and\ \citenamefont {Zhang}(2024)}]{Chen2024}%
  \BibitemOpen
  \bibfield  {author} {\bibinfo {author} {\bibfnamefont {Y.-Q.}\ \bibnamefont {Chen}}\ and\ \bibinfo {author} {\bibfnamefont {S.-X.}\ \bibnamefont {Zhang}},\ }\bibfield  {title} {Effective temperature in approximate quantum many-body states,\ }\href@noop {} {\bibfield  {journal} {\bibinfo  {journal} {arXiv:2411.18921}\ } (\bibinfo {year} {2024})}\BibitemShut {NoStop}%
\bibitem [{\citenamefont {Östlund}\ and\ \citenamefont {Rommer}(1995)}]{stlund1995}%
  \BibitemOpen
  \bibfield  {author} {\bibinfo {author} {\bibfnamefont {S.}~\bibnamefont {Östlund}}\ and\ \bibinfo {author} {\bibfnamefont {S.}~\bibnamefont {Rommer}},\ }\bibfield  {title} {Thermodynamic limit of density matrix renormalization,\ }\href {\doibase 10.1103/PhysRevLett.75.3537} {\bibfield  {journal} {\bibinfo  {journal} {Physical Review Letters}\ }\textbf {\bibinfo {volume} {75}},\ \bibinfo {pages} {3537} (\bibinfo {year} {1995})}\BibitemShut {NoStop}%
\bibitem [{\citenamefont {Pirvu}\ \emph {et~al.}(2012)\citenamefont {Pirvu}, \citenamefont {Haegeman},\ and\ \citenamefont {Verstraete}}]{Pirvu2012}%
  \BibitemOpen
  \bibfield  {author} {\bibinfo {author} {\bibfnamefont {B.}~\bibnamefont {Pirvu}}, \bibinfo {author} {\bibfnamefont {J.}~\bibnamefont {Haegeman}}, \ and\ \bibinfo {author} {\bibfnamefont {F.}~\bibnamefont {Verstraete}},\ }\bibfield  {title} {Matrix product state based algorithm for determining dispersion relations of quantum spin chains with periodic boundary conditions,\ }\href {\doibase 10.1103/PhysRevB.85.035130} {\bibfield  {journal} {\bibinfo  {journal} {Physical Review B}\ }\textbf {\bibinfo {volume} {85}},\ \bibinfo {pages} {035130} (\bibinfo {year} {2012})}\BibitemShut {NoStop}%
\bibitem [{\citenamefont {Haegeman}\ \emph {et~al.}(2012)\citenamefont {Haegeman}, \citenamefont {Pirvu}, \citenamefont {Weir}, \citenamefont {Cirac}, \citenamefont {Osborne}, \citenamefont {Verschelde},\ and\ \citenamefont {Verstraete}}]{Haegeman2012}%
  \BibitemOpen
  \bibfield  {author} {\bibinfo {author} {\bibfnamefont {J.}~\bibnamefont {Haegeman}}, \bibinfo {author} {\bibfnamefont {B.}~\bibnamefont {Pirvu}}, \bibinfo {author} {\bibfnamefont {D.~J.}\ \bibnamefont {Weir}}, \bibinfo {author} {\bibfnamefont {J.~I.}\ \bibnamefont {Cirac}}, \bibinfo {author} {\bibfnamefont {T.~J.}\ \bibnamefont {Osborne}}, \bibinfo {author} {\bibfnamefont {H.}~\bibnamefont {Verschelde}}, \ and\ \bibinfo {author} {\bibfnamefont {F.}~\bibnamefont {Verstraete}},\ }\bibfield  {title} {Variational matrix product ansatz for dispersion relations,\ }\href {\doibase 10.1103/PhysRevB.85.100408} {\bibfield  {journal} {\bibinfo  {journal} {Physical Review B}\ }\textbf {\bibinfo {volume} {85}},\ \bibinfo {pages} {100408} (\bibinfo {year} {2012})}\BibitemShut {NoStop}%
\bibitem [{\citenamefont {Vanderstraeten}\ \emph {et~al.}(2018)\citenamefont {Vanderstraeten}, \citenamefont {Damme}, \citenamefont {Büchler},\ and\ \citenamefont {Verstraete}}]{Vanderstraeten2018}%
  \BibitemOpen
  \bibfield  {author} {\bibinfo {author} {\bibfnamefont {L.}~\bibnamefont {Vanderstraeten}}, \bibinfo {author} {\bibfnamefont {M.~V.}\ \bibnamefont {Damme}}, \bibinfo {author} {\bibfnamefont {H.~P.}\ \bibnamefont {Büchler}}, \ and\ \bibinfo {author} {\bibfnamefont {F.}~\bibnamefont {Verstraete}},\ }\bibfield  {title} {Quasiparticles in quantum spin chains with long-range interactions,\ }\href {\doibase 10.1103/PhysRevLett.121.090603} {\bibfield  {journal} {\bibinfo  {journal} {Physical Review Letters}\ }\textbf {\bibinfo {volume} {121}},\ \bibinfo {pages} {090603} (\bibinfo {year} {2018})}\BibitemShut {NoStop}%
\bibitem [{\citenamefont {Zou}\ \emph {et~al.}(2018)\citenamefont {Zou}, \citenamefont {Milsted},\ and\ \citenamefont {Vidal}}]{Zou2018}%
  \BibitemOpen
  \bibfield  {author} {\bibinfo {author} {\bibfnamefont {Y.}~\bibnamefont {Zou}}, \bibinfo {author} {\bibfnamefont {A.}~\bibnamefont {Milsted}}, \ and\ \bibinfo {author} {\bibfnamefont {G.}~\bibnamefont {Vidal}},\ }\bibfield  {title} {Conformal data and renormalization group flow in critical quantum spin chains using periodic uniform matrix product states,\ }\href {\doibase 10.1103/PhysRevLett.121.230402} {\bibfield  {journal} {\bibinfo  {journal} {Physical Review Letters}\ }\textbf {\bibinfo {volume} {121}},\ \bibinfo {pages} {230402} (\bibinfo {year} {2018})}\BibitemShut {NoStop}%
\bibitem [{\citenamefont {Vanderstraeten}\ \emph {et~al.}(2019)\citenamefont {Vanderstraeten}, \citenamefont {Haegeman},\ and\ \citenamefont {Verstraete}}]{Vanderstraeten2019}%
  \BibitemOpen
  \bibfield  {author} {\bibinfo {author} {\bibfnamefont {L.}~\bibnamefont {Vanderstraeten}}, \bibinfo {author} {\bibfnamefont {J.}~\bibnamefont {Haegeman}}, \ and\ \bibinfo {author} {\bibfnamefont {F.}~\bibnamefont {Verstraete}},\ }\bibfield  {title} {Simulating excitation spectra with projected entangled-pair states,\ }\href {\doibase 10.1103/PhysRevB.99.165121} {\bibfield  {journal} {\bibinfo  {journal} {Physical Review B}\ }\textbf {\bibinfo {volume} {99}},\ \bibinfo {pages} {165121} (\bibinfo {year} {2019})}\BibitemShut {NoStop}%
\bibitem [{\citenamefont {Ponsioen}\ and\ \citenamefont {Corboz}(2020)}]{Ponsioen2020}%
  \BibitemOpen
  \bibfield  {author} {\bibinfo {author} {\bibfnamefont {B.}~\bibnamefont {Ponsioen}}\ and\ \bibinfo {author} {\bibfnamefont {P.}~\bibnamefont {Corboz}},\ }\bibfield  {title} {Excitations with projected entangled pair states using the corner transfer matrix method,\ }\href {\doibase 10.1103/PhysRevB.101.195109} {\bibfield  {journal} {\bibinfo  {journal} {Physical Review B}\ }\textbf {\bibinfo {volume} {101}},\ \bibinfo {pages} {195109} (\bibinfo {year} {2020})}\BibitemShut {NoStop}%
\bibitem [{\citenamefont {Tu}\ \emph {et~al.}(2021)\citenamefont {Tu}, \citenamefont {Wu}, \citenamefont {Schuch}, \citenamefont {Kawashima},\ and\ \citenamefont {Chen}}]{Tu2021}%
  \BibitemOpen
  \bibfield  {author} {\bibinfo {author} {\bibfnamefont {W.-L.}\ \bibnamefont {Tu}}, \bibinfo {author} {\bibfnamefont {H.-K.}\ \bibnamefont {Wu}}, \bibinfo {author} {\bibfnamefont {N.}~\bibnamefont {Schuch}}, \bibinfo {author} {\bibfnamefont {N.}~\bibnamefont {Kawashima}}, \ and\ \bibinfo {author} {\bibfnamefont {J.-Y.}\ \bibnamefont {Chen}},\ }\bibfield  {title} {Generating function for tensor network diagrammatic summation,\ }\href {\doibase 10.1103/PhysRevB.103.205155} {\bibfield  {journal} {\bibinfo  {journal} {Physical Review B}\ }\textbf {\bibinfo {volume} {103}},\ \bibinfo {pages} {205155} (\bibinfo {year} {2021})}\BibitemShut {NoStop}%
\bibitem [{\citenamefont {White}(1992)}]{White1992}%
  \BibitemOpen
  \bibfield  {author} {\bibinfo {author} {\bibfnamefont {S.~R.}\ \bibnamefont {White}},\ }\bibfield  {title} {Density matrix formulation for quantum renormalization groups,\ }\href {\doibase 10.1103/PhysRevLett.69.2863} {\bibfield  {journal} {\bibinfo  {journal} {Physical Review Letters}\ }\textbf {\bibinfo {volume} {69}},\ \bibinfo {pages} {2863} (\bibinfo {year} {1992})}\BibitemShut {NoStop}%
\bibitem [{\citenamefont {Bañuls}\ \emph {et~al.}(2013)\citenamefont {Bañuls}, \citenamefont {Cichy}, \citenamefont {Cirac},\ and\ \citenamefont {Jansen}}]{Bauls2013}%
  \BibitemOpen
  \bibfield  {author} {\bibinfo {author} {\bibfnamefont {M.}~\bibnamefont {Bañuls}}, \bibinfo {author} {\bibfnamefont {K.}~\bibnamefont {Cichy}}, \bibinfo {author} {\bibfnamefont {J.}~\bibnamefont {Cirac}}, \ and\ \bibinfo {author} {\bibfnamefont {K.}~\bibnamefont {Jansen}},\ }\bibfield  {title} {The mass spectrum of the schwinger model with matrix product states,\ }\href {\doibase 10.1007/JHEP11(2013)158} {\bibfield  {journal} {\bibinfo  {journal} {Journal of High Energy Physics}\ }\textbf {\bibinfo {volume} {2013}},\ \bibinfo {pages} {158} (\bibinfo {year} {2013})}\BibitemShut {NoStop}%
\bibitem [{\citenamefont {Choo}\ \emph {et~al.}(2018)\citenamefont {Choo}, \citenamefont {Carleo}, \citenamefont {Regnault},\ and\ \citenamefont {Neupert}}]{Choo2018}%
  \BibitemOpen
  \bibfield  {author} {\bibinfo {author} {\bibfnamefont {K.}~\bibnamefont {Choo}}, \bibinfo {author} {\bibfnamefont {G.}~\bibnamefont {Carleo}}, \bibinfo {author} {\bibfnamefont {N.}~\bibnamefont {Regnault}}, \ and\ \bibinfo {author} {\bibfnamefont {T.}~\bibnamefont {Neupert}},\ }\bibfield  {title} {Symmetries and many-body excitations with neural-network quantum states,\ }\href {\doibase 10.1103/PhysRevLett.121.167204} {\bibfield  {journal} {\bibinfo  {journal} {Physical Review Letters}\ }\textbf {\bibinfo {volume} {121}},\ \bibinfo {pages} {167204} (\bibinfo {year} {2018})}\BibitemShut {NoStop}%
\bibitem [{\citenamefont {Jones}\ \emph {et~al.}(2019)\citenamefont {Jones}, \citenamefont {Endo}, \citenamefont {McArdle}, \citenamefont {Yuan},\ and\ \citenamefont {Benjamin}}]{Jones2019}%
  \BibitemOpen
  \bibfield  {author} {\bibinfo {author} {\bibfnamefont {T.}~\bibnamefont {Jones}}, \bibinfo {author} {\bibfnamefont {S.}~\bibnamefont {Endo}}, \bibinfo {author} {\bibfnamefont {S.}~\bibnamefont {McArdle}}, \bibinfo {author} {\bibfnamefont {X.}~\bibnamefont {Yuan}}, \ and\ \bibinfo {author} {\bibfnamefont {S.~C.}\ \bibnamefont {Benjamin}},\ }\bibfield  {title} {Variational quantum algorithms for discovering hamiltonian spectra,\ }\href {\doibase 10.1103/PhysRevA.99.062304} {\bibfield  {journal} {\bibinfo  {journal} {Physical Review A}\ }\textbf {\bibinfo {volume} {99}},\ \bibinfo {pages} {062304} (\bibinfo {year} {2019})}\BibitemShut {NoStop}%
\bibitem [{\citenamefont {Pathak}\ \emph {et~al.}(2021)\citenamefont {Pathak}, \citenamefont {Busemeyer}, \citenamefont {Rodrigues},\ and\ \citenamefont {Wagner}}]{Pathak2021}%
  \BibitemOpen
  \bibfield  {author} {\bibinfo {author} {\bibfnamefont {S.}~\bibnamefont {Pathak}}, \bibinfo {author} {\bibfnamefont {B.}~\bibnamefont {Busemeyer}}, \bibinfo {author} {\bibfnamefont {J.~N.~B.}\ \bibnamefont {Rodrigues}}, \ and\ \bibinfo {author} {\bibfnamefont {L.~K.}\ \bibnamefont {Wagner}},\ }\bibfield  {title} {Excited states in variational monte carlo using a penalty method,\ }\href {https://pubs.aip.org/aip/jcp/article/154/3/034101/199876/Excited-states-in-variational-Monte-Carlo-using-a} {\bibfield  {journal} {\bibinfo  {journal} {The Journal of Chemical Physics}\ }\textbf {\bibinfo {volume} {154}} (\bibinfo {year} {2021})}\BibitemShut {NoStop}%
\bibitem [{\citenamefont {Entwistle}\ \emph {et~al.}(2023)\citenamefont {Entwistle}, \citenamefont {Schätzle}, \citenamefont {Erdman}, \citenamefont {Hermann},\ and\ \citenamefont {Noé}}]{Entwistle2023}%
  \BibitemOpen
  \bibfield  {author} {\bibinfo {author} {\bibfnamefont {M.~T.}\ \bibnamefont {Entwistle}}, \bibinfo {author} {\bibfnamefont {Z.}~\bibnamefont {Schätzle}}, \bibinfo {author} {\bibfnamefont {P.~A.}\ \bibnamefont {Erdman}}, \bibinfo {author} {\bibfnamefont {J.}~\bibnamefont {Hermann}}, \ and\ \bibinfo {author} {\bibfnamefont {F.}~\bibnamefont {Noé}},\ }\bibfield  {title} {Electronic excited states in deep variational monte carlo,\ }\href {\doibase 10.1038/s41467-022-35534-5} {\bibfield  {journal} {\bibinfo  {journal} {Nature Communications}\ }\textbf {\bibinfo {volume} {14}},\ \bibinfo {pages} {274} (\bibinfo {year} {2023})}\BibitemShut {NoStop}%
\bibitem [{\citenamefont {Larsson}(2025)}]{Larsson2025}%
  \BibitemOpen
  \bibfield  {author} {\bibinfo {author} {\bibfnamefont {H.~R.}\ \bibnamefont {Larsson}},\ }\bibfield  {title} {Benchmarking vibrational spectra: 5000 accurate eigenstates of acetonitrile using tree tensor network states,\ }\href {\doibase 10.1021/acs.jpclett.5c00782} {\bibfield  {journal} {\bibinfo  {journal} {The Journal of Physical Chemistry Letters}\ }\textbf {\bibinfo {volume} {16}},\ \bibinfo {pages} {3991} (\bibinfo {year} {2025})}\BibitemShut {NoStop}%
\bibitem [{\citenamefont {Higgott}\ \emph {et~al.}(2019)\citenamefont {Higgott}, \citenamefont {Wang},\ and\ \citenamefont {Brierley}}]{Higgott2019}%
  \BibitemOpen
  \bibfield  {author} {\bibinfo {author} {\bibfnamefont {O.}~\bibnamefont {Higgott}}, \bibinfo {author} {\bibfnamefont {D.}~\bibnamefont {Wang}}, \ and\ \bibinfo {author} {\bibfnamefont {S.}~\bibnamefont {Brierley}},\ }\bibfield  {title} {Variational quantum computation of excited states,\ }\href {\doibase 10.22331/q-2019-07-01-156} {\bibfield  {journal} {\bibinfo  {journal} {Quantum}\ }\textbf {\bibinfo {volume} {3}},\ \bibinfo {pages} {156} (\bibinfo {year} {2019})}\BibitemShut {NoStop}%
\bibitem [{\citenamefont {Wheeler}\ \emph {et~al.}(2024)\citenamefont {Wheeler}, \citenamefont {Kleiner},\ and\ \citenamefont {Wagner}}]{Wheeler2024}%
  \BibitemOpen
  \bibfield  {author} {\bibinfo {author} {\bibfnamefont {W.~A.}\ \bibnamefont {Wheeler}}, \bibinfo {author} {\bibfnamefont {K.~G.}\ \bibnamefont {Kleiner}}, \ and\ \bibinfo {author} {\bibfnamefont {L.~K.}\ \bibnamefont {Wagner}},\ }\bibfield  {title} {Ensemble variational monte carlo for optimization of correlated excited state wave functions,\ }\href {\doibase 10.1088/2516-1075/ad38f8} {\bibfield  {journal} {\bibinfo  {journal} {Electronic Structure}\ }\textbf {\bibinfo {volume} {6}},\ \bibinfo {pages} {025001} (\bibinfo {year} {2024})}\BibitemShut {NoStop}%
\bibitem [{\citenamefont {Quiroga}\ \emph {et~al.}(2025)\citenamefont {Quiroga}, \citenamefont {Han},\ and\ \citenamefont {Kyrillidis}}]{Quiroga2025}%
  \BibitemOpen
  \bibfield  {author} {\bibinfo {author} {\bibfnamefont {D.}~\bibnamefont {Quiroga}}, \bibinfo {author} {\bibfnamefont {J.}~\bibnamefont {Han}}, \ and\ \bibinfo {author} {\bibfnamefont {A.}~\bibnamefont {Kyrillidis}},\ }\bibfield  {title} {Quantum eigengame for excited state calculation,\ }\href@noop {} {\bibfield  {journal} {\bibinfo  {journal} {arXiv:2503.13644}\ } (\bibinfo {year} {2025})}\BibitemShut {NoStop}%
\bibitem [{\citenamefont {Nakanishi}\ \emph {et~al.}(2019)\citenamefont {Nakanishi}, \citenamefont {Mitarai},\ and\ \citenamefont {Fujii}}]{Nakanishi2019}%
  \BibitemOpen
  \bibfield  {author} {\bibinfo {author} {\bibfnamefont {K.~M.}\ \bibnamefont {Nakanishi}}, \bibinfo {author} {\bibfnamefont {K.}~\bibnamefont {Mitarai}}, \ and\ \bibinfo {author} {\bibfnamefont {K.}~\bibnamefont {Fujii}},\ }\bibfield  {title} {Subspace-search variational quantum eigensolver for excited states,\ }\href {\doibase 10.1103/PhysRevResearch.1.033062} {\bibfield  {journal} {\bibinfo  {journal} {Physical Review Research}\ }\textbf {\bibinfo {volume} {1}},\ \bibinfo {pages} {033062} (\bibinfo {year} {2019})}\BibitemShut {NoStop}%
\bibitem [{\citenamefont {LaRose}\ \emph {et~al.}(2019)\citenamefont {LaRose}, \citenamefont {Tikku}, \citenamefont {Étude O’Neel-Judy}, \citenamefont {Cincio},\ and\ \citenamefont {Coles}}]{LaRose2019}%
  \BibitemOpen
  \bibfield  {author} {\bibinfo {author} {\bibfnamefont {R.}~\bibnamefont {LaRose}}, \bibinfo {author} {\bibfnamefont {A.}~\bibnamefont {Tikku}}, \bibinfo {author} {\bibnamefont {Étude O’Neel-Judy}}, \bibinfo {author} {\bibfnamefont {L.}~\bibnamefont {Cincio}}, \ and\ \bibinfo {author} {\bibfnamefont {P.~J.}\ \bibnamefont {Coles}},\ }\bibfield  {title} {Variational quantum state diagonalization,\ }\href {\doibase 10.1038/s41534-019-0167-6} {\bibfield  {journal} {\bibinfo  {journal} {npj Quantum Information}\ }\textbf {\bibinfo {volume} {5}},\ \bibinfo {pages} {57} (\bibinfo {year} {2019})}\BibitemShut {NoStop}%
\bibitem [{\citenamefont {Parrish}\ and\ \citenamefont {McMahon}(2019)}]{Parrish2019}%
  \BibitemOpen
  \bibfield  {author} {\bibinfo {author} {\bibfnamefont {R.~M.}\ \bibnamefont {Parrish}}\ and\ \bibinfo {author} {\bibfnamefont {P.~L.}\ \bibnamefont {McMahon}},\ }\bibfield  {title} {Quantum filter diagonalization: Quantum eigendecomposition without full quantum phase estimation,\ }\href@noop {} {\bibfield  {journal} {\bibinfo  {journal} {arXiv:1909.08925}\ } (\bibinfo {year} {2019})}\BibitemShut {NoStop}%
\bibitem [{\citenamefont {Li}\ \emph {et~al.}(2024)\citenamefont {Li}, \citenamefont {Zhou}, \citenamefont {Xu}, \citenamefont {Chi}, \citenamefont {Guo}, \citenamefont {Liu}, \citenamefont {Liao},\ and\ \citenamefont {Xiang}}]{Li2024}%
  \BibitemOpen
  \bibfield  {author} {\bibinfo {author} {\bibfnamefont {X.}~\bibnamefont {Li}}, \bibinfo {author} {\bibfnamefont {Z.}~\bibnamefont {Zhou}}, \bibinfo {author} {\bibfnamefont {G.}~\bibnamefont {Xu}}, \bibinfo {author} {\bibfnamefont {R.}~\bibnamefont {Chi}}, \bibinfo {author} {\bibfnamefont {Y.}~\bibnamefont {Guo}}, \bibinfo {author} {\bibfnamefont {T.}~\bibnamefont {Liu}}, \bibinfo {author} {\bibfnamefont {H.}~\bibnamefont {Liao}}, \ and\ \bibinfo {author} {\bibfnamefont {T.}~\bibnamefont {Xiang}},\ }\bibfield  {title} {Accurate determination of low-energy eigenspectra with multitarget matrix product states,\ }\href {\doibase 10.1103/PhysRevB.109.045115} {\bibfield  {journal} {\bibinfo  {journal} {Physical Review B}\ }\textbf {\bibinfo {volume} {109}},\ \bibinfo {pages} {045115} (\bibinfo {year} {2024})}\BibitemShut {NoStop}%
\bibitem [{\citenamefont {Zhang}\ \emph {et~al.}(2024)\citenamefont {Zhang}, \citenamefont {Wang},\ and\ \citenamefont {Wang}}]{zhang2024neural}%
  \BibitemOpen
  \bibfield  {author} {\bibinfo {author} {\bibfnamefont {Q.}~\bibnamefont {Zhang}}, \bibinfo {author} {\bibfnamefont {R.-S.}\ \bibnamefont {Wang}}, \ and\ \bibinfo {author} {\bibfnamefont {L.}~\bibnamefont {Wang}},\ }\bibfield  {title} {Neural canonical transformations for vibrational spectra of molecules,\ }\href {\doibase 10.1063/5.0209255} {\bibfield  {journal} {\bibinfo  {journal} {The Journal of Chemical Physics}\ }\textbf {\bibinfo {volume} {161}} (\bibinfo {year} {2024}),\ 10.1063/5.0209255}\BibitemShut {NoStop}%
\bibitem [{\citenamefont {Umrigar}\ \emph {et~al.}(1988)\citenamefont {Umrigar}, \citenamefont {Wilson},\ and\ \citenamefont {Wilkins}}]{Umrigar1988}%
  \BibitemOpen
  \bibfield  {author} {\bibinfo {author} {\bibfnamefont {C.~J.}\ \bibnamefont {Umrigar}}, \bibinfo {author} {\bibfnamefont {K.~G.}\ \bibnamefont {Wilson}}, \ and\ \bibinfo {author} {\bibfnamefont {J.~W.}\ \bibnamefont {Wilkins}},\ }\bibfield  {title} {Optimized trial wave functions for quantum monte carlo calculations,\ }\href {\doibase 10.1103/PhysRevLett.60.1719} {\bibfield  {journal} {\bibinfo  {journal} {Physical Review Letters}\ }\textbf {\bibinfo {volume} {60}},\ \bibinfo {pages} {1719} (\bibinfo {year} {1988})}\BibitemShut {NoStop}%
\bibitem [{\citenamefont {Siringo}\ and\ \citenamefont {Marotta}(2005)}]{Siringo2005}%
  \BibitemOpen
  \bibfield  {author} {\bibinfo {author} {\bibfnamefont {F.}~\bibnamefont {Siringo}}\ and\ \bibinfo {author} {\bibfnamefont {L.}~\bibnamefont {Marotta}},\ }\bibfield  {title} {A variational method from the variance of energy,\ }\href {\doibase 10.1140/epjc/s2005-02358-x} {\bibfield  {journal} {\bibinfo  {journal} {The European Physical Journal C}\ }\textbf {\bibinfo {volume} {44}},\ \bibinfo {pages} {293} (\bibinfo {year} {2005})}\BibitemShut {NoStop}%
\bibitem [{\citenamefont {Umrigar}\ and\ \citenamefont {Filippi}(2005)}]{Umrigar2005}%
  \BibitemOpen
  \bibfield  {author} {\bibinfo {author} {\bibfnamefont {C.~J.}\ \bibnamefont {Umrigar}}\ and\ \bibinfo {author} {\bibfnamefont {C.}~\bibnamefont {Filippi}},\ }\bibfield  {title} {Energy and variance optimization of many-body wave functions,\ }\href {\doibase 10.1103/PhysRevLett.94.150201} {\bibfield  {journal} {\bibinfo  {journal} {Physical Review Letters}\ }\textbf {\bibinfo {volume} {94}},\ \bibinfo {pages} {150201} (\bibinfo {year} {2005})}\BibitemShut {NoStop}%
\bibitem [{\citenamefont {Pollmann}\ \emph {et~al.}(2016)\citenamefont {Pollmann}, \citenamefont {Khemani}, \citenamefont {Cirac},\ and\ \citenamefont {Sondhi}}]{Pollmann2016}%
  \BibitemOpen
  \bibfield  {author} {\bibinfo {author} {\bibfnamefont {F.}~\bibnamefont {Pollmann}}, \bibinfo {author} {\bibfnamefont {V.}~\bibnamefont {Khemani}}, \bibinfo {author} {\bibfnamefont {J.~I.}\ \bibnamefont {Cirac}}, \ and\ \bibinfo {author} {\bibfnamefont {S.~L.}\ \bibnamefont {Sondhi}},\ }\bibfield  {title} {Efficient variational diagonalization of fully many-body localized hamiltonians,\ }\href {\doibase 10.1103/PhysRevB.94.041116} {\bibfield  {journal} {\bibinfo  {journal} {Physical Review B}\ }\textbf {\bibinfo {volume} {94}},\ \bibinfo {pages} {041116} (\bibinfo {year} {2016})}\BibitemShut {NoStop}%
\bibitem [{\citenamefont {Vicentini}\ \emph {et~al.}(2019)\citenamefont {Vicentini}, \citenamefont {Biella}, \citenamefont {Regnault},\ and\ \citenamefont {Ciuti}}]{Vicentini2019}%
  \BibitemOpen
  \bibfield  {author} {\bibinfo {author} {\bibfnamefont {F.}~\bibnamefont {Vicentini}}, \bibinfo {author} {\bibfnamefont {A.}~\bibnamefont {Biella}}, \bibinfo {author} {\bibfnamefont {N.}~\bibnamefont {Regnault}}, \ and\ \bibinfo {author} {\bibfnamefont {C.}~\bibnamefont {Ciuti}},\ }\bibfield  {title} {Variational neural-network ansatz for steady states in open quantum systems,\ }\href {\doibase 10.1103/PhysRevLett.122.250503} {\bibfield  {journal} {\bibinfo  {journal} {Physical Review Letters}\ }\textbf {\bibinfo {volume} {122}},\ \bibinfo {pages} {250503} (\bibinfo {year} {2019})}\BibitemShut {NoStop}%
\bibitem [{\citenamefont {Zhang}\ \emph {et~al.}(2021)\citenamefont {Zhang}, \citenamefont {Gomes}, \citenamefont {Yao}, \citenamefont {Orth},\ and\ \citenamefont {Iadecola}}]{Zhang2021}%
  \BibitemOpen
  \bibfield  {author} {\bibinfo {author} {\bibfnamefont {F.}~\bibnamefont {Zhang}}, \bibinfo {author} {\bibfnamefont {N.}~\bibnamefont {Gomes}}, \bibinfo {author} {\bibfnamefont {Y.}~\bibnamefont {Yao}}, \bibinfo {author} {\bibfnamefont {P.~P.}\ \bibnamefont {Orth}}, \ and\ \bibinfo {author} {\bibfnamefont {T.}~\bibnamefont {Iadecola}},\ }\bibfield  {title} {Adaptive variational quantum eigensolvers for highly excited states,\ }\href {\doibase 10.1103/PhysRevB.104.075159} {\bibfield  {journal} {\bibinfo  {journal} {Physical Review B}\ }\textbf {\bibinfo {volume} {104}},\ \bibinfo {pages} {075159} (\bibinfo {year} {2021})}\BibitemShut {NoStop}%
\bibitem [{\citenamefont {Liu}\ \emph {et~al.}(2023)\citenamefont {Liu}, \citenamefont {Zhang}, \citenamefont {Hsieh}, \citenamefont {Zhang},\ and\ \citenamefont {Yao}}]{Liu2021c}%
  \BibitemOpen
  \bibfield  {author} {\bibinfo {author} {\bibfnamefont {S.}~\bibnamefont {Liu}}, \bibinfo {author} {\bibfnamefont {S.-X.}\ \bibnamefont {Zhang}}, \bibinfo {author} {\bibfnamefont {C.-Y.}\ \bibnamefont {Hsieh}}, \bibinfo {author} {\bibfnamefont {S.}~\bibnamefont {Zhang}}, \ and\ \bibinfo {author} {\bibfnamefont {H.}~\bibnamefont {Yao}},\ }\bibfield  {title} {Probing many-body localization by excited-state variational quantum eigensolver,\ }\href {\doibase 10.1103/PhysRevB.107.024204} {\bibfield  {journal} {\bibinfo  {journal} {Physical Review B}\ }\textbf {\bibinfo {volume} {107}},\ \bibinfo {pages} {024204} (\bibinfo {year} {2023})}\BibitemShut {NoStop}%
\bibitem [{\citenamefont {Zhang}\ \emph {et~al.}(2022)\citenamefont {Zhang}, \citenamefont {Chen}, \citenamefont {Yuan},\ and\ \citenamefont {Yin}}]{Zhang2022}%
  \BibitemOpen
  \bibfield  {author} {\bibinfo {author} {\bibfnamefont {D.-B.}\ \bibnamefont {Zhang}}, \bibinfo {author} {\bibfnamefont {B.-L.}\ \bibnamefont {Chen}}, \bibinfo {author} {\bibfnamefont {Z.-H.}\ \bibnamefont {Yuan}}, \ and\ \bibinfo {author} {\bibfnamefont {T.}~\bibnamefont {Yin}},\ }\bibfield  {title} {Variational quantum eigensolvers by variance minimization,\ }\href {\doibase 10.1088/1674-1056/ac8a8d} {\bibfield  {journal} {\bibinfo  {journal} {Chinese Physics B}\ }\textbf {\bibinfo {volume} {31}},\ \bibinfo {pages} {120301} (\bibinfo {year} {2022})}\BibitemShut {NoStop}%
\bibitem [{\citenamefont {Wang}\ and\ \citenamefont {Zunger}(1994)}]{Wang1994}%
  \BibitemOpen
  \bibfield  {author} {\bibinfo {author} {\bibfnamefont {L.-W.}\ \bibnamefont {Wang}}\ and\ \bibinfo {author} {\bibfnamefont {A.}~\bibnamefont {Zunger}},\ }\bibfield  {title} {Solving schrödinger’s equation around a desired energy: Application to silicon quantum dots,\ }\href {\doibase 10.1063/1.466486} {\bibfield  {journal} {\bibinfo  {journal} {The Journal of Chemical Physics}\ }\textbf {\bibinfo {volume} {100}},\ \bibinfo {pages} {2394} (\bibinfo {year} {1994})}\BibitemShut {NoStop}%
\bibitem [{\citenamefont {McClean}\ \emph {et~al.}(2016)\citenamefont {McClean}, \citenamefont {Romero}, \citenamefont {Babbush},\ and\ \citenamefont {Aspuru-Guzik}}]{McClean2016}%
  \BibitemOpen
  \bibfield  {author} {\bibinfo {author} {\bibfnamefont {J.~R.}\ \bibnamefont {McClean}}, \bibinfo {author} {\bibfnamefont {J.}~\bibnamefont {Romero}}, \bibinfo {author} {\bibfnamefont {R.}~\bibnamefont {Babbush}}, \ and\ \bibinfo {author} {\bibfnamefont {A.}~\bibnamefont {Aspuru-Guzik}},\ }\bibfield  {title} {The theory of variational hybrid quantum-classical algorithms,\ }\href {\doibase 10.1088/1367-2630/18/2/023023} {\bibfield  {journal} {\bibinfo  {journal} {New Journal of Physics}\ }\textbf {\bibinfo {volume} {18}},\ \bibinfo {pages} {023023} (\bibinfo {year} {2016})}\BibitemShut {NoStop}%
\bibitem [{\citenamefont {Santagati}\ \emph {et~al.}(2018)\citenamefont {Santagati}, \citenamefont {Wang}, \citenamefont {Gentile}, \citenamefont {Paesani}, \citenamefont {Wiebe}, \citenamefont {McClean}, \citenamefont {Morley-Short}, \citenamefont {Shadbolt}, \citenamefont {Bonneau}, \citenamefont {Silverstone}, \citenamefont {Tew}, \citenamefont {Zhou}, \citenamefont {O’Brien},\ and\ \citenamefont {Thompson}}]{Santagati2018}%
  \BibitemOpen
  \bibfield  {author} {\bibinfo {author} {\bibfnamefont {R.}~\bibnamefont {Santagati}}, \bibinfo {author} {\bibfnamefont {J.}~\bibnamefont {Wang}}, \bibinfo {author} {\bibfnamefont {A.~A.}\ \bibnamefont {Gentile}}, \bibinfo {author} {\bibfnamefont {S.}~\bibnamefont {Paesani}}, \bibinfo {author} {\bibfnamefont {N.}~\bibnamefont {Wiebe}}, \bibinfo {author} {\bibfnamefont {J.~R.}\ \bibnamefont {McClean}}, \bibinfo {author} {\bibfnamefont {S.}~\bibnamefont {Morley-Short}}, \bibinfo {author} {\bibfnamefont {P.~J.}\ \bibnamefont {Shadbolt}}, \bibinfo {author} {\bibfnamefont {D.}~\bibnamefont {Bonneau}}, \bibinfo {author} {\bibfnamefont {J.~W.}\ \bibnamefont {Silverstone}}, \bibinfo {author} {\bibfnamefont {D.~P.}\ \bibnamefont {Tew}}, \bibinfo {author} {\bibfnamefont {X.}~\bibnamefont {Zhou}}, \bibinfo {author} {\bibfnamefont {J.~L.}\ \bibnamefont {O’Brien}}, \ and\ \bibinfo {author} {\bibfnamefont {M.~G.}\ \bibnamefont {Thompson}},\ }\bibfield  {title} {Witnessing eigenstates for quantum simulation of
  hamiltonian spectra,\ }\href {\doibase 10.1126/sciadv.aap9646} {\bibfield  {journal} {\bibinfo  {journal} {Science Advances}\ }\textbf {\bibinfo {volume} {4}},\ \bibinfo {pages} {1} (\bibinfo {year} {2018})}\BibitemShut {NoStop}%
\bibitem [{\citenamefont {Wang}\ and\ \citenamefont {Mazziotti}(2023)}]{Wang2023}%
  \BibitemOpen
  \bibfield  {author} {\bibinfo {author} {\bibfnamefont {Y.}~\bibnamefont {Wang}}\ and\ \bibinfo {author} {\bibfnamefont {D.~A.}\ \bibnamefont {Mazziotti}},\ }\bibfield  {title} {Electronic excited states from a variance-based contracted quantum eigensolver,\ }\href {\doibase 10.1103/PhysRevA.108.022814} {\bibfield  {journal} {\bibinfo  {journal} {Physical Review A}\ }\textbf {\bibinfo {volume} {108}},\ \bibinfo {pages} {022814} (\bibinfo {year} {2023})}\BibitemShut {NoStop}%
\bibitem [{\citenamefont {Cenedese}\ \emph {et~al.}(2024)\citenamefont {Cenedese}, \citenamefont {Bondani}, \citenamefont {Andreanov}, \citenamefont {Carrega}, \citenamefont {Benenti},\ and\ \citenamefont {Rosa}}]{Cenedese2024}%
  \BibitemOpen
  \bibfield  {author} {\bibinfo {author} {\bibfnamefont {G.}~\bibnamefont {Cenedese}}, \bibinfo {author} {\bibfnamefont {M.}~\bibnamefont {Bondani}}, \bibinfo {author} {\bibfnamefont {A.}~\bibnamefont {Andreanov}}, \bibinfo {author} {\bibfnamefont {M.}~\bibnamefont {Carrega}}, \bibinfo {author} {\bibfnamefont {G.}~\bibnamefont {Benenti}}, \ and\ \bibinfo {author} {\bibfnamefont {D.}~\bibnamefont {Rosa}},\ }\bibfield  {title} {Shallow quantum circuits are robust hunters for quantum many-body scars,\ }\href@noop {} {\bibfield  {journal} {\bibinfo  {journal} {arXiv:2401.09279}\ } (\bibinfo {year} {2024})}\BibitemShut {NoStop}%
\bibitem [{\citenamefont {Edelman}\ \emph {et~al.}(1998)\citenamefont {Edelman}, \citenamefont {Arias},\ and\ \citenamefont {Smith}}]{Edelman1998}%
  \BibitemOpen
  \bibfield  {author} {\bibinfo {author} {\bibfnamefont {A.}~\bibnamefont {Edelman}}, \bibinfo {author} {\bibfnamefont {T.~A.}\ \bibnamefont {Arias}}, \ and\ \bibinfo {author} {\bibfnamefont {S.~T.}\ \bibnamefont {Smith}},\ }\bibfield  {title} {The geometry of algorithms with orthogonality constraints,\ }\href {\doibase 10.1137/S0895479895290954} {\bibfield  {journal} {\bibinfo  {journal} {SIAM Journal on Matrix Analysis and Applications}\ }\textbf {\bibinfo {volume} {20}},\ \bibinfo {pages} {303} (\bibinfo {year} {1998})}\BibitemShut {NoStop}%
\bibitem [{\citenamefont {Pfau}\ \emph {et~al.}(2019)\citenamefont {Pfau}, \citenamefont {Petersen}, \citenamefont {Agarwal}, \citenamefont {Barrett},\ and\ \citenamefont {Stachenfeld}}]{Pfau2019}%
  \BibitemOpen
  \bibfield  {author} {\bibinfo {author} {\bibfnamefont {D.}~\bibnamefont {Pfau}}, \bibinfo {author} {\bibfnamefont {S.}~\bibnamefont {Petersen}}, \bibinfo {author} {\bibfnamefont {A.}~\bibnamefont {Agarwal}}, \bibinfo {author} {\bibfnamefont {D.~G.~T.}\ \bibnamefont {Barrett}}, \ and\ \bibinfo {author} {\bibfnamefont {K.~L.}\ \bibnamefont {Stachenfeld}},\ }\bibfield  {title} {Spectral inference networks: Unifying deep and spectral learning,\ }\href {https://arxiv.org/abs/1806.02215} {\bibfield  {journal} {\bibinfo  {journal} {ICLR 2019}\ } (\bibinfo {year} {2019})}\BibitemShut {NoStop}%
\bibitem [{\citenamefont {Pfau}\ \emph {et~al.}(2024)\citenamefont {Pfau}, \citenamefont {Axelrod}, \citenamefont {Sutterud}, \citenamefont {von Glehn},\ and\ \citenamefont {Spencer}}]{Pfau2024}%
  \BibitemOpen
  \bibfield  {author} {\bibinfo {author} {\bibfnamefont {D.}~\bibnamefont {Pfau}}, \bibinfo {author} {\bibfnamefont {S.}~\bibnamefont {Axelrod}}, \bibinfo {author} {\bibfnamefont {H.}~\bibnamefont {Sutterud}}, \bibinfo {author} {\bibfnamefont {I.}~\bibnamefont {von Glehn}}, \ and\ \bibinfo {author} {\bibfnamefont {J.~S.}\ \bibnamefont {Spencer}},\ }\bibfield  {title} {Accurate computation of quantum excited states with neural networks,\ }\href {https://www.science.org/doi/10.1126/science.adn0137} {\bibfield  {journal} {\bibinfo  {journal} {Science}\ }\textbf {\bibinfo {volume} {385}} (\bibinfo {year} {2024})}\BibitemShut {NoStop}%
\bibitem [{\citenamefont {Schollw{\"{o}}ck}(2011)}]{Schollwock2011}%
  \BibitemOpen
  \bibfield  {author} {\bibinfo {author} {\bibfnamefont {U.}~\bibnamefont {Schollw{\"{o}}ck}},\ }\bibfield  {title} {{The density-matrix renormalization group in the age of matrix product states},\ }\href {\doibase 10.1016/j.aop.2010.09.012} {\bibfield  {journal} {\bibinfo  {journal} {Ann. Phys.}\ }\textbf {\bibinfo {volume} {326}},\ \bibinfo {pages} {96} (\bibinfo {year} {2011})}\BibitemShut {NoStop}%
\bibitem [{\citenamefont {Stoudenmire}\ and\ \citenamefont {White}(2012)}]{Stoudenmire2012}%
  \BibitemOpen
  \bibfield  {author} {\bibinfo {author} {\bibfnamefont {E.}~\bibnamefont {Stoudenmire}}\ and\ \bibinfo {author} {\bibfnamefont {S.~R.}\ \bibnamefont {White}},\ }\bibfield  {title} {{Studying Two-Dimensional Systems with the Density Matrix Renormalization Group},\ }\href {\doibase 10.1146/annurev-conmatphys-020911-125018} {\bibfield  {journal} {\bibinfo  {journal} {Annu. Rev. Condens. Matter Phys.}\ }\textbf {\bibinfo {volume} {3}},\ \bibinfo {pages} {111} (\bibinfo {year} {2012})}\BibitemShut {NoStop}%
\bibitem [{\citenamefont {Khoromskij}(2011)}]{Khoromskij2011}%
  \BibitemOpen
  \bibfield  {author} {\bibinfo {author} {\bibfnamefont {B.~N.}\ \bibnamefont {Khoromskij}},\ }\bibfield  {title} {O(dlogn)-quantics approximation of n-d tensors in high-dimensional numerical modeling,\ }\href {\doibase 10.1007/s00365-011-9131-1} {\bibfield  {journal} {\bibinfo  {journal} {Constructive Approximation}\ }\textbf {\bibinfo {volume} {34}},\ \bibinfo {pages} {257} (\bibinfo {year} {2011})}\BibitemShut {NoStop}%
\bibitem [{\citenamefont {Fernández}\ \emph {et~al.}(2025)\citenamefont {Fernández}, \citenamefont {Ritter}, \citenamefont {Jeannin}, \citenamefont {Li}, \citenamefont {Kloss}, \citenamefont {Louvet}, \citenamefont {Terasaki}, \citenamefont {Parcollet}, \citenamefont {von Delft}, \citenamefont {Shinaoka},\ and\ \citenamefont {Waintal}}]{NezFernndez2025}%
  \BibitemOpen
  \bibfield  {author} {\bibinfo {author} {\bibfnamefont {Y.~N.}\ \bibnamefont {Fernández}}, \bibinfo {author} {\bibfnamefont {M.~K.}\ \bibnamefont {Ritter}}, \bibinfo {author} {\bibfnamefont {M.}~\bibnamefont {Jeannin}}, \bibinfo {author} {\bibfnamefont {J.-W.}\ \bibnamefont {Li}}, \bibinfo {author} {\bibfnamefont {T.}~\bibnamefont {Kloss}}, \bibinfo {author} {\bibfnamefont {T.}~\bibnamefont {Louvet}}, \bibinfo {author} {\bibfnamefont {S.}~\bibnamefont {Terasaki}}, \bibinfo {author} {\bibfnamefont {O.}~\bibnamefont {Parcollet}}, \bibinfo {author} {\bibfnamefont {J.}~\bibnamefont {von Delft}}, \bibinfo {author} {\bibfnamefont {H.}~\bibnamefont {Shinaoka}}, \ and\ \bibinfo {author} {\bibfnamefont {X.}~\bibnamefont {Waintal}},\ }\bibfield  {title} {Learning tensor networks with tensor cross interpolation: New algorithms and libraries,\ }\href {\doibase 10.21468/SciPostPhys.18.3.104} {\bibfield  {journal} {\bibinfo  {journal} {SciPost Physics}\ }\textbf {\bibinfo {volume} {18}},\ \bibinfo {pages} {104} (\bibinfo
  {year} {2025})}\BibitemShut {NoStop}%
\bibitem [{\citenamefont {Lubasch}\ \emph {et~al.}(2018)\citenamefont {Lubasch}, \citenamefont {Moinier},\ and\ \citenamefont {Jaksch}}]{Lubasch2018}%
  \BibitemOpen
  \bibfield  {author} {\bibinfo {author} {\bibfnamefont {M.}~\bibnamefont {Lubasch}}, \bibinfo {author} {\bibfnamefont {P.}~\bibnamefont {Moinier}}, \ and\ \bibinfo {author} {\bibfnamefont {D.}~\bibnamefont {Jaksch}},\ }\bibfield  {title} {Multigrid renormalization,\ }\href {\doibase 10.1016/j.jcp.2018.06.065} {\bibfield  {journal} {\bibinfo  {journal} {Journal of Computational Physics}\ }\textbf {\bibinfo {volume} {372}},\ \bibinfo {pages} {587} (\bibinfo {year} {2018})}\BibitemShut {NoStop}%
\bibitem [{\citenamefont {García-Ripoll}(2021)}]{Garca-Ripoll2021}%
  \BibitemOpen
  \bibfield  {author} {\bibinfo {author} {\bibfnamefont {J.~J.}\ \bibnamefont {García-Ripoll}},\ }\bibfield  {title} {Quantum-inspired algorithms for multivariate analysis: from interpolation to partial differential equations,\ }\href {\doibase 10.22331/q-2021-04-15-431} {\bibfield  {journal} {\bibinfo  {journal} {Quantum}\ }\textbf {\bibinfo {volume} {5}},\ \bibinfo {pages} {431} (\bibinfo {year} {2021})}\BibitemShut {NoStop}%
\bibitem [{\citenamefont {Ye}\ and\ \citenamefont {Loureiro}(2022)}]{Ye2022}%
  \BibitemOpen
  \bibfield  {author} {\bibinfo {author} {\bibfnamefont {E.}~\bibnamefont {Ye}}\ and\ \bibinfo {author} {\bibfnamefont {N.~F.~G.}\ \bibnamefont {Loureiro}},\ }\bibfield  {title} {Quantum-inspired method for solving the vlasov-poisson equations,\ }\href {\doibase 10.1103/PhysRevE.106.035208} {\bibfield  {journal} {\bibinfo  {journal} {Physical Review E}\ }\textbf {\bibinfo {volume} {106}},\ \bibinfo {pages} {035208} (\bibinfo {year} {2022})}\BibitemShut {NoStop}%
\bibitem [{\citenamefont {Gourianov}\ \emph {et~al.}(2022)\citenamefont {Gourianov}, \citenamefont {Lubasch}, \citenamefont {Dolgov}, \citenamefont {van~den Berg}, \citenamefont {Babaee}, \citenamefont {Givi}, \citenamefont {Kiffner},\ and\ \citenamefont {Jaksch}}]{Gourianov2022}%
  \BibitemOpen
  \bibfield  {author} {\bibinfo {author} {\bibfnamefont {N.}~\bibnamefont {Gourianov}}, \bibinfo {author} {\bibfnamefont {M.}~\bibnamefont {Lubasch}}, \bibinfo {author} {\bibfnamefont {S.}~\bibnamefont {Dolgov}}, \bibinfo {author} {\bibfnamefont {Q.~Y.}\ \bibnamefont {van~den Berg}}, \bibinfo {author} {\bibfnamefont {H.}~\bibnamefont {Babaee}}, \bibinfo {author} {\bibfnamefont {P.}~\bibnamefont {Givi}}, \bibinfo {author} {\bibfnamefont {M.}~\bibnamefont {Kiffner}}, \ and\ \bibinfo {author} {\bibfnamefont {D.}~\bibnamefont {Jaksch}},\ }\bibfield  {title} {A quantum-inspired approach to exploit turbulence structures,\ }\href {\doibase 10.1038/s43588-021-00181-1} {\bibfield  {journal} {\bibinfo  {journal} {Nature Computational Science}\ }\textbf {\bibinfo {volume} {2}},\ \bibinfo {pages} {30} (\bibinfo {year} {2022})}\BibitemShut {NoStop}%
\bibitem [{\citenamefont {Shinaoka}\ \emph {et~al.}(2023)\citenamefont {Shinaoka}, \citenamefont {Wallerberger}, \citenamefont {Murakami}, \citenamefont {Nogaki}, \citenamefont {Sakurai}, \citenamefont {Werner},\ and\ \citenamefont {Kauch}}]{Shinaoka2023}%
  \BibitemOpen
  \bibfield  {author} {\bibinfo {author} {\bibfnamefont {H.}~\bibnamefont {Shinaoka}}, \bibinfo {author} {\bibfnamefont {M.}~\bibnamefont {Wallerberger}}, \bibinfo {author} {\bibfnamefont {Y.}~\bibnamefont {Murakami}}, \bibinfo {author} {\bibfnamefont {K.}~\bibnamefont {Nogaki}}, \bibinfo {author} {\bibfnamefont {R.}~\bibnamefont {Sakurai}}, \bibinfo {author} {\bibfnamefont {P.}~\bibnamefont {Werner}}, \ and\ \bibinfo {author} {\bibfnamefont {A.}~\bibnamefont {Kauch}},\ }\bibfield  {title} {Multiscale space-time ansatz for correlation functions of quantum systems based on quantics tensor trains,\ }\href {\doibase 10.1103/PhysRevX.13.021015} {\bibfield  {journal} {\bibinfo  {journal} {Physical Review X}\ }\textbf {\bibinfo {volume} {13}},\ \bibinfo {pages} {021015} (\bibinfo {year} {2023})}\BibitemShut {NoStop}%
\bibitem [{\citenamefont {Ritter}\ \emph {et~al.}(2024)\citenamefont {Ritter}, \citenamefont {Fernández}, \citenamefont {Wallerberger}, \citenamefont {von Delft}, \citenamefont {Shinaoka},\ and\ \citenamefont {Waintal}}]{Ritter2024}%
  \BibitemOpen
  \bibfield  {author} {\bibinfo {author} {\bibfnamefont {M.~K.}\ \bibnamefont {Ritter}}, \bibinfo {author} {\bibfnamefont {Y.~N.}\ \bibnamefont {Fernández}}, \bibinfo {author} {\bibfnamefont {M.}~\bibnamefont {Wallerberger}}, \bibinfo {author} {\bibfnamefont {J.}~\bibnamefont {von Delft}}, \bibinfo {author} {\bibfnamefont {H.}~\bibnamefont {Shinaoka}}, \ and\ \bibinfo {author} {\bibfnamefont {X.}~\bibnamefont {Waintal}},\ }\bibfield  {title} {Quantics tensor cross interpolation for high-resolution parsimonious representations of multivariate functions,\ }\href {\doibase 10.1103/PhysRevLett.132.056501} {\bibfield  {journal} {\bibinfo  {journal} {Physical Review Letters}\ }\textbf {\bibinfo {volume} {132}},\ \bibinfo {pages} {056501} (\bibinfo {year} {2024})}\BibitemShut {NoStop}%
\bibitem [{\citenamefont {Peruzzo}\ \emph {et~al.}(2014)\citenamefont {Peruzzo}, \citenamefont {McClean}, \citenamefont {Shadbolt}, \citenamefont {Yung}, \citenamefont {Zhou}, \citenamefont {Love}, \citenamefont {Aspuru-Guzik},\ and\ \citenamefont {O’Brien}}]{Peruzzo2014}%
  \BibitemOpen
  \bibfield  {author} {\bibinfo {author} {\bibfnamefont {A.}~\bibnamefont {Peruzzo}}, \bibinfo {author} {\bibfnamefont {J.}~\bibnamefont {McClean}}, \bibinfo {author} {\bibfnamefont {P.}~\bibnamefont {Shadbolt}}, \bibinfo {author} {\bibfnamefont {M.-H.}\ \bibnamefont {Yung}}, \bibinfo {author} {\bibfnamefont {X.-Q.}\ \bibnamefont {Zhou}}, \bibinfo {author} {\bibfnamefont {P.~J.}\ \bibnamefont {Love}}, \bibinfo {author} {\bibfnamefont {A.}~\bibnamefont {Aspuru-Guzik}}, \ and\ \bibinfo {author} {\bibfnamefont {J.~L.}\ \bibnamefont {O’Brien}},\ }\bibfield  {title} {A variational eigenvalue solver on a photonic quantum processor,\ }\href {\doibase 10.1038/ncomms5213} {\bibfield  {journal} {\bibinfo  {journal} {Nature Communications}\ }\textbf {\bibinfo {volume} {5}},\ \bibinfo {pages} {4213} (\bibinfo {year} {2014})}\BibitemShut {NoStop}%
\bibitem [{\citenamefont {Bharti}\ \emph {et~al.}(2022)\citenamefont {Bharti}, \citenamefont {Cervera-Lierta}, \citenamefont {Kyaw}, \citenamefont {Haug}, \citenamefont {Alperin-Lea}, \citenamefont {Anand}, \citenamefont {Degroote}, \citenamefont {Heimonen}, \citenamefont {Kottmann}, \citenamefont {Menke}, \citenamefont {Mok}, \citenamefont {Sim}, \citenamefont {Kwek},\ and\ \citenamefont {Aspuru-Guzik}}]{Bharti2021}%
  \BibitemOpen
  \bibfield  {author} {\bibinfo {author} {\bibfnamefont {K.}~\bibnamefont {Bharti}}, \bibinfo {author} {\bibfnamefont {A.}~\bibnamefont {Cervera-Lierta}}, \bibinfo {author} {\bibfnamefont {T.~H.}\ \bibnamefont {Kyaw}}, \bibinfo {author} {\bibfnamefont {T.}~\bibnamefont {Haug}}, \bibinfo {author} {\bibfnamefont {S.}~\bibnamefont {Alperin-Lea}}, \bibinfo {author} {\bibfnamefont {A.}~\bibnamefont {Anand}}, \bibinfo {author} {\bibfnamefont {M.}~\bibnamefont {Degroote}}, \bibinfo {author} {\bibfnamefont {H.}~\bibnamefont {Heimonen}}, \bibinfo {author} {\bibfnamefont {J.~S.}\ \bibnamefont {Kottmann}}, \bibinfo {author} {\bibfnamefont {T.}~\bibnamefont {Menke}}, \bibinfo {author} {\bibfnamefont {W.-K.}\ \bibnamefont {Mok}}, \bibinfo {author} {\bibfnamefont {S.}~\bibnamefont {Sim}}, \bibinfo {author} {\bibfnamefont {L.-C.}\ \bibnamefont {Kwek}}, \ and\ \bibinfo {author} {\bibfnamefont {A.}~\bibnamefont {Aspuru-Guzik}},\ }\bibfield  {title} {Noisy intermediate-scale quantum algorithms,\ }\href {\doibase
  10.1103/RevModPhys.94.015004} {\bibfield  {journal} {\bibinfo  {journal} {Reviews of Modern Physics}\ }\textbf {\bibinfo {volume} {94}},\ \bibinfo {pages} {015004} (\bibinfo {year} {2022})}\BibitemShut {NoStop}%
\bibitem [{\citenamefont {Cerezo}\ \emph {et~al.}(2021)\citenamefont {Cerezo}, \citenamefont {Arrasmith}, \citenamefont {Babbush}, \citenamefont {Benjamin}, \citenamefont {Endo}, \citenamefont {Fujii}, \citenamefont {McClean}, \citenamefont {Mitarai}, \citenamefont {Yuan}, \citenamefont {Cincio},\ and\ \citenamefont {Coles}}]{Cerezo2020b}%
  \BibitemOpen
  \bibfield  {author} {\bibinfo {author} {\bibfnamefont {M.}~\bibnamefont {Cerezo}}, \bibinfo {author} {\bibfnamefont {A.}~\bibnamefont {Arrasmith}}, \bibinfo {author} {\bibfnamefont {R.}~\bibnamefont {Babbush}}, \bibinfo {author} {\bibfnamefont {S.~C.}\ \bibnamefont {Benjamin}}, \bibinfo {author} {\bibfnamefont {S.}~\bibnamefont {Endo}}, \bibinfo {author} {\bibfnamefont {K.}~\bibnamefont {Fujii}}, \bibinfo {author} {\bibfnamefont {J.~R.}\ \bibnamefont {McClean}}, \bibinfo {author} {\bibfnamefont {K.}~\bibnamefont {Mitarai}}, \bibinfo {author} {\bibfnamefont {X.}~\bibnamefont {Yuan}}, \bibinfo {author} {\bibfnamefont {L.}~\bibnamefont {Cincio}}, \ and\ \bibinfo {author} {\bibfnamefont {P.~J.}\ \bibnamefont {Coles}},\ }\bibfield  {title} {Variational quantum algorithms,\ }\href {\doibase 10.1038/s42254-021-00348-9} {\bibfield  {journal} {\bibinfo  {journal} {Nature Reviews Physics}\ }\textbf {\bibinfo {volume} {3}},\ \bibinfo {pages} {625} (\bibinfo {year} {2021})}\BibitemShut {NoStop}%
\bibitem [{\citenamefont {Tilly}\ \emph {et~al.}(2022)\citenamefont {Tilly}, \citenamefont {Chen}, \citenamefont {Cao}, \citenamefont {Picozzi}, \citenamefont {Setia}, \citenamefont {Li}, \citenamefont {Grant}, \citenamefont {Wossnig}, \citenamefont {Rungger}, \citenamefont {Booth},\ and\ \citenamefont {Tennyson}}]{Tilly2022}%
  \BibitemOpen
  \bibfield  {author} {\bibinfo {author} {\bibfnamefont {J.}~\bibnamefont {Tilly}}, \bibinfo {author} {\bibfnamefont {H.}~\bibnamefont {Chen}}, \bibinfo {author} {\bibfnamefont {S.}~\bibnamefont {Cao}}, \bibinfo {author} {\bibfnamefont {D.}~\bibnamefont {Picozzi}}, \bibinfo {author} {\bibfnamefont {K.}~\bibnamefont {Setia}}, \bibinfo {author} {\bibfnamefont {Y.}~\bibnamefont {Li}}, \bibinfo {author} {\bibfnamefont {E.}~\bibnamefont {Grant}}, \bibinfo {author} {\bibfnamefont {L.}~\bibnamefont {Wossnig}}, \bibinfo {author} {\bibfnamefont {I.}~\bibnamefont {Rungger}}, \bibinfo {author} {\bibfnamefont {G.~H.}\ \bibnamefont {Booth}}, \ and\ \bibinfo {author} {\bibfnamefont {J.}~\bibnamefont {Tennyson}},\ }\bibfield  {title} {The variational quantum eigensolver: A review of methods and best practices,\ }\href {\doibase 10.1016/j.physrep.2022.08.003} {\bibfield  {journal} {\bibinfo  {journal} {Physics Reports}\ }\textbf {\bibinfo {volume} {986}},\ \bibinfo {pages} {1} (\bibinfo {year} {2022})}\BibitemShut {NoStop}%
\bibitem [{Note1()}]{Note1}%
  \BibitemOpen
  \bibinfo {note} {See supplemental materials for details}\BibitemShut {NoStop}%
\bibitem [{\citenamefont {Liao}\ \emph {et~al.}(2019)\citenamefont {Liao}, \citenamefont {Liu}, \citenamefont {Wang},\ and\ \citenamefont {Xiang}}]{Liao2019}%
  \BibitemOpen
  \bibfield  {author} {\bibinfo {author} {\bibfnamefont {H.-J.}\ \bibnamefont {Liao}}, \bibinfo {author} {\bibfnamefont {J.-G.}\ \bibnamefont {Liu}}, \bibinfo {author} {\bibfnamefont {L.}~\bibnamefont {Wang}}, \ and\ \bibinfo {author} {\bibfnamefont {T.}~\bibnamefont {Xiang}},\ }\bibfield  {title} {Differentiable programming tensor networks,\ }\href {\doibase 10.1103/physrevx.9.031041} {\bibfield  {journal} {\bibinfo  {journal} {Physical Review X}\ }\textbf {\bibinfo {volume} {9}},\ \bibinfo {pages} {31041} (\bibinfo {year} {2019})}\BibitemShut {NoStop}%
\bibitem [{\citenamefont {Zhang}\ \emph {et~al.}(2023{\natexlab{a}})\citenamefont {Zhang}, \citenamefont {Wan},\ and\ \citenamefont {Yao}}]{Zhang2019b}%
  \BibitemOpen
  \bibfield  {author} {\bibinfo {author} {\bibfnamefont {S.-X.}\ \bibnamefont {Zhang}}, \bibinfo {author} {\bibfnamefont {Z.-Q.}\ \bibnamefont {Wan}}, \ and\ \bibinfo {author} {\bibfnamefont {H.}~\bibnamefont {Yao}},\ }\bibfield  {title} {Automatic differentiable monte carlo: Theory and application,\ }\href {\doibase 10.1103/PhysRevResearch.5.033041} {\bibfield  {journal} {\bibinfo  {journal} {Physical Review Research}\ }\textbf {\bibinfo {volume} {5}},\ \bibinfo {pages} {033041} (\bibinfo {year} {2023}{\natexlab{a}})}\BibitemShut {NoStop}%
\bibitem [{\citenamefont {Zhang}\ \emph {et~al.}(2023{\natexlab{b}})\citenamefont {Zhang}, \citenamefont {Allcock}, \citenamefont {Wan}, \citenamefont {Liu}, \citenamefont {Sun}, \citenamefont {Yu}, \citenamefont {Yang}, \citenamefont {Qiu}, \citenamefont {Ye}, \citenamefont {Chen}, \citenamefont {Lee}, \citenamefont {Zheng}, \citenamefont {Jian}, \citenamefont {Yao}, \citenamefont {Hsieh},\ and\ \citenamefont {Zhang}}]{Zhang2022_z}%
  \BibitemOpen
  \bibfield  {author} {\bibinfo {author} {\bibfnamefont {S.-X.}\ \bibnamefont {Zhang}}, \bibinfo {author} {\bibfnamefont {J.}~\bibnamefont {Allcock}}, \bibinfo {author} {\bibfnamefont {Z.-Q.}\ \bibnamefont {Wan}}, \bibinfo {author} {\bibfnamefont {S.}~\bibnamefont {Liu}}, \bibinfo {author} {\bibfnamefont {J.}~\bibnamefont {Sun}}, \bibinfo {author} {\bibfnamefont {H.}~\bibnamefont {Yu}}, \bibinfo {author} {\bibfnamefont {X.-H.}\ \bibnamefont {Yang}}, \bibinfo {author} {\bibfnamefont {J.}~\bibnamefont {Qiu}}, \bibinfo {author} {\bibfnamefont {Z.}~\bibnamefont {Ye}}, \bibinfo {author} {\bibfnamefont {Y.-Q.}\ \bibnamefont {Chen}}, \bibinfo {author} {\bibfnamefont {C.-K.}\ \bibnamefont {Lee}}, \bibinfo {author} {\bibfnamefont {Y.-C.}\ \bibnamefont {Zheng}}, \bibinfo {author} {\bibfnamefont {S.-K.}\ \bibnamefont {Jian}}, \bibinfo {author} {\bibfnamefont {H.}~\bibnamefont {Yao}}, \bibinfo {author} {\bibfnamefont {C.-Y.}\ \bibnamefont {Hsieh}}, \ and\ \bibinfo {author} {\bibfnamefont {S.}~\bibnamefont {Zhang}},\
  }\bibfield  {title} {{TensorCircuit: a Quantum Software Framework for the NISQ Era},\ }\href {https://quantum-journal.org/papers/q-2023-02-02-912/} {\bibfield  {journal} {\bibinfo  {journal} {Quantum}\ }\textbf {\bibinfo {volume} {7}},\ \bibinfo {pages} {912} (\bibinfo {year} {2023}{\natexlab{b}})}\BibitemShut {NoStop}%
\bibitem [{\citenamefont {Verstraete}\ and\ \citenamefont {Cirac}(2004{\natexlab{a}})}]{Verstraete2004}%
  \BibitemOpen
  \bibfield  {author} {\bibinfo {author} {\bibfnamefont {F.}~\bibnamefont {Verstraete}}\ and\ \bibinfo {author} {\bibfnamefont {J.~I.}\ \bibnamefont {Cirac}},\ }\bibfield  {title} {{Renormalization algorithms for Quantum-Many Body Systems in two and higher dimensions},\ }\href {http://arxiv.org/abs/cond-mat/0407066} {\bibfield  {journal} {\bibinfo  {journal} {arXiv:cond-mat/0407066}\ } (\bibinfo {year} {2004}{\natexlab{a}})}\BibitemShut {NoStop}%
\bibitem [{\citenamefont {Verstraete}\ and\ \citenamefont {Cirac}(2004{\natexlab{b}})}]{Verstraete2004a}%
  \BibitemOpen
  \bibfield  {author} {\bibinfo {author} {\bibfnamefont {F.}~\bibnamefont {Verstraete}}\ and\ \bibinfo {author} {\bibfnamefont {J.~I.}\ \bibnamefont {Cirac}},\ }\bibfield  {title} {{Valence-bond states for quantum computation},\ }\href {\doibase 10.1103/PhysRevA.70.060302} {\bibfield  {journal} {\bibinfo  {journal} {Phys. Rev. A}\ }\textbf {\bibinfo {volume} {70}},\ \bibinfo {pages} {060302} (\bibinfo {year} {2004}{\natexlab{b}})}\BibitemShut {NoStop}%
\bibitem [{\citenamefont {Vogt}\ \emph {et~al.}(2019)\citenamefont {Vogt}, \citenamefont {Sage},\ and\ \citenamefont {Kjaergaard}}]{Vogt2019}%
  \BibitemOpen
  \bibfield  {author} {\bibinfo {author} {\bibfnamefont {E.}~\bibnamefont {Vogt}}, \bibinfo {author} {\bibfnamefont {D.~S.}\ \bibnamefont {Sage}}, \ and\ \bibinfo {author} {\bibfnamefont {H.~G.}\ \bibnamefont {Kjaergaard}},\ }\bibfield  {title} {Accuracy of xh-stretching intensities with the deng–fan potential,\ }\href {\doibase 10.1080/00268976.2018.1521529} {\bibfield  {journal} {\bibinfo  {journal} {Molecular Physics}\ }\textbf {\bibinfo {volume} {117}},\ \bibinfo {pages} {1629} (\bibinfo {year} {2019})}\BibitemShut {NoStop}%
\bibitem [{\citenamefont {Vogt}\ \emph {et~al.}(2025)\citenamefont {Vogt}, \citenamefont {Álvaro Fernández~Corral}, \citenamefont {Saleh},\ and\ \citenamefont {Yachmenev}}]{Vogt2025}%
  \BibitemOpen
  \bibfield  {author} {\bibinfo {author} {\bibfnamefont {E.}~\bibnamefont {Vogt}}, \bibinfo {author} {\bibnamefont {Álvaro Fernández~Corral}}, \bibinfo {author} {\bibfnamefont {Y.}~\bibnamefont {Saleh}}, \ and\ \bibinfo {author} {\bibfnamefont {A.}~\bibnamefont {Yachmenev}},\ }\bibfield  {title} {Transferability of vibrational normalizing-flow coordinates: A pathway towards intrinsic coordinates,\ }\href@noop {} {\bibfield  {journal} {\bibinfo  {journal} {arXiv:2502.15750}\ } (\bibinfo {year} {2025})}\BibitemShut {NoStop}%
\bibitem [{\citenamefont {Jolly}\ \emph {et~al.}(2023)\citenamefont {Jolly}, \citenamefont {Fernández},\ and\ \citenamefont {Waintal}}]{Jolly2023}%
  \BibitemOpen
  \bibfield  {author} {\bibinfo {author} {\bibfnamefont {N.}~\bibnamefont {Jolly}}, \bibinfo {author} {\bibfnamefont {Y.~N.}\ \bibnamefont {Fernández}}, \ and\ \bibinfo {author} {\bibfnamefont {X.}~\bibnamefont {Waintal}},\ }\bibfield  {title} {Tensorized orbitals for computational chemistry,\ }\href@noop {} {\bibfield  {journal} {\bibinfo  {journal} {arXiv:2308.03508}\ } (\bibinfo {year} {2023})}\BibitemShut {NoStop}%
\bibitem [{\citenamefont {Huggins}\ \emph {et~al.}(2020)\citenamefont {Huggins}, \citenamefont {Lee}, \citenamefont {Baek}, \citenamefont {O’Gorman},\ and\ \citenamefont {Whaley}}]{Huggins2020}%
  \BibitemOpen
  \bibfield  {author} {\bibinfo {author} {\bibfnamefont {W.~J.}\ \bibnamefont {Huggins}}, \bibinfo {author} {\bibfnamefont {J.}~\bibnamefont {Lee}}, \bibinfo {author} {\bibfnamefont {U.}~\bibnamefont {Baek}}, \bibinfo {author} {\bibfnamefont {B.}~\bibnamefont {O’Gorman}}, \ and\ \bibinfo {author} {\bibfnamefont {K.~B.}\ \bibnamefont {Whaley}},\ }\bibfield  {title} {A non-orthogonal variational quantum eigensolver,\ }\href {\doibase 10.1088/1367-2630/ab867b} {\bibfield  {journal} {\bibinfo  {journal} {New Journal of Physics}\ }\textbf {\bibinfo {volume} {22}},\ \bibinfo {pages} {073009} (\bibinfo {year} {2020})}\BibitemShut {NoStop}%
\bibitem [{\citenamefont {Carleo}\ and\ \citenamefont {Troyer}(2017)}]{Carleo2017}%
  \BibitemOpen
  \bibfield  {author} {\bibinfo {author} {\bibfnamefont {G.}~\bibnamefont {Carleo}}\ and\ \bibinfo {author} {\bibfnamefont {M.}~\bibnamefont {Troyer}},\ }\bibfield  {title} {Solving the quantum many-body problem with artificial neural networks,\ }\href {\doibase 10.1126/science.aag2302} {\bibfield  {journal} {\bibinfo  {journal} {Science}\ }\textbf {\bibinfo {volume} {355}},\ \bibinfo {pages} {602} (\bibinfo {year} {2017})}\BibitemShut {NoStop}%
\bibitem [{\citenamefont {Hinze}(1973)}]{Hinze1973}%
  \BibitemOpen
  \bibfield  {author} {\bibinfo {author} {\bibfnamefont {J.}~\bibnamefont {Hinze}},\ }\bibfield  {title} {Mc-scf. i. the multi-configuration self-consistent-field method,\ }\href {\doibase 10.1063/1.1680022} {\bibfield  {journal} {\bibinfo  {journal} {The Journal of Chemical Physics}\ }\textbf {\bibinfo {volume} {59}},\ \bibinfo {pages} {6424} (\bibinfo {year} {1973})}\BibitemShut {NoStop}%
\bibitem [{\citenamefont {Diffenderfer}\ and\ \citenamefont {Yarkony}(1982)}]{Diffenderfer1982}%
  \BibitemOpen
  \bibfield  {author} {\bibinfo {author} {\bibfnamefont {R.~N.}\ \bibnamefont {Diffenderfer}}\ and\ \bibinfo {author} {\bibfnamefont {D.~R.}\ \bibnamefont {Yarkony}},\ }\bibfield  {title} {Use of the state-averaged mcscf procedure: application to radiative transitions in magnesium oxide,\ }\href {\doibase 10.1021/j100223a010} {\bibfield  {journal} {\bibinfo  {journal} {The Journal of Physical Chemistry}\ }\textbf {\bibinfo {volume} {86}},\ \bibinfo {pages} {5098} (\bibinfo {year} {1982})}\BibitemShut {NoStop}%
\bibitem [{\citenamefont {McClean}\ \emph {et~al.}(2017)\citenamefont {McClean}, \citenamefont {Kimchi-Schwartz}, \citenamefont {Carter},\ and\ \citenamefont {de~Jong}}]{McClean2017}%
  \BibitemOpen
  \bibfield  {author} {\bibinfo {author} {\bibfnamefont {J.~R.}\ \bibnamefont {McClean}}, \bibinfo {author} {\bibfnamefont {M.~E.}\ \bibnamefont {Kimchi-Schwartz}}, \bibinfo {author} {\bibfnamefont {J.}~\bibnamefont {Carter}}, \ and\ \bibinfo {author} {\bibfnamefont {W.~A.}\ \bibnamefont {de~Jong}},\ }\bibfield  {title} {Hybrid quantum-classical hierarchy for mitigation of decoherence and determination of excited states,\ }\href {\doibase 10.1103/PhysRevA.95.042308} {\bibfield  {journal} {\bibinfo  {journal} {Physical Review A}\ }\textbf {\bibinfo {volume} {95}},\ \bibinfo {pages} {042308} (\bibinfo {year} {2017})}\BibitemShut {NoStop}%
\bibitem [{\citenamefont {Motta}\ \emph {et~al.}(2020)\citenamefont {Motta}, \citenamefont {Sun}, \citenamefont {Tan}, \citenamefont {O’Rourke}, \citenamefont {Ye}, \citenamefont {Minnich}, \citenamefont {Brandão},\ and\ \citenamefont {lic Lic~Chan}}]{Motta2020}%
  \BibitemOpen
  \bibfield  {author} {\bibinfo {author} {\bibfnamefont {M.}~\bibnamefont {Motta}}, \bibinfo {author} {\bibfnamefont {C.}~\bibnamefont {Sun}}, \bibinfo {author} {\bibfnamefont {A.~T.~K.}\ \bibnamefont {Tan}}, \bibinfo {author} {\bibfnamefont {M.~J.}\ \bibnamefont {O’Rourke}}, \bibinfo {author} {\bibfnamefont {E.}~\bibnamefont {Ye}}, \bibinfo {author} {\bibfnamefont {A.~J.}\ \bibnamefont {Minnich}}, \bibinfo {author} {\bibfnamefont {F.~G. S.~L.}\ \bibnamefont {Brandão}}, \ and\ \bibinfo {author} {\bibfnamefont {G.~K.}\ \bibnamefont {lic Lic~Chan}},\ }\bibfield  {title} {Determining eigenstates and thermal states on a quantum computer using quantum imaginary time evolution,\ }\href {\doibase 10.1038/s41567-019-0704-4} {\bibfield  {journal} {\bibinfo  {journal} {Nature Physics}\ }\textbf {\bibinfo {volume} {16}},\ \bibinfo {pages} {205} (\bibinfo {year} {2020})}\BibitemShut {NoStop}%
\bibitem [{\citenamefont {Baek}\ \emph {et~al.}(2023)\citenamefont {Baek}, \citenamefont {Hait}, \citenamefont {Shee}, \citenamefont {Leimkuhler}, \citenamefont {Huggins}, \citenamefont {Stetina}, \citenamefont {Head-Gordon},\ and\ \citenamefont {Whaley}}]{Baek2023}%
  \BibitemOpen
  \bibfield  {author} {\bibinfo {author} {\bibfnamefont {U.}~\bibnamefont {Baek}}, \bibinfo {author} {\bibfnamefont {D.}~\bibnamefont {Hait}}, \bibinfo {author} {\bibfnamefont {J.}~\bibnamefont {Shee}}, \bibinfo {author} {\bibfnamefont {O.}~\bibnamefont {Leimkuhler}}, \bibinfo {author} {\bibfnamefont {W.~J.}\ \bibnamefont {Huggins}}, \bibinfo {author} {\bibfnamefont {T.~F.}\ \bibnamefont {Stetina}}, \bibinfo {author} {\bibfnamefont {M.}~\bibnamefont {Head-Gordon}}, \ and\ \bibinfo {author} {\bibfnamefont {K.~B.}\ \bibnamefont {Whaley}},\ }\bibfield  {title} {Say no to optimization: A nonorthogonal quantum eigensolver,\ }\href {\doibase 10.1103/PRXQuantum.4.030307} {\bibfield  {journal} {\bibinfo  {journal} {PRX Quantum}\ }\textbf {\bibinfo {volume} {4}},\ \bibinfo {pages} {030307} (\bibinfo {year} {2023})}\BibitemShut {NoStop}%
\bibitem [{\citenamefont {Giuliani}\ \emph {et~al.}(2025)\citenamefont {Giuliani}, \citenamefont {Nys}, \citenamefont {Martinazzo}, \citenamefont {Carleo},\ and\ \citenamefont {Rossi}}]{Giuliani2025}%
  \BibitemOpen
  \bibfield  {author} {\bibinfo {author} {\bibfnamefont {C.}~\bibnamefont {Giuliani}}, \bibinfo {author} {\bibfnamefont {J.}~\bibnamefont {Nys}}, \bibinfo {author} {\bibfnamefont {R.}~\bibnamefont {Martinazzo}}, \bibinfo {author} {\bibfnamefont {G.}~\bibnamefont {Carleo}}, \ and\ \bibinfo {author} {\bibfnamefont {R.}~\bibnamefont {Rossi}},\ }\bibfield  {title} {Precise quantum chemistry calculations with few slater determinants,\ }\href@noop {} {\bibfield  {journal} {\bibinfo  {journal} {arXiv:2503.14502}\ } (\bibinfo {year} {2025})}\BibitemShut {NoStop}%
\bibitem [{\citenamefont {Ordejón}\ \emph {et~al.}(1993)\citenamefont {Ordejón}, \citenamefont {Drabold}, \citenamefont {Grumbach},\ and\ \citenamefont {Martin}}]{Ordejn1993}%
  \BibitemOpen
  \bibfield  {author} {\bibinfo {author} {\bibfnamefont {P.}~\bibnamefont {Ordejón}}, \bibinfo {author} {\bibfnamefont {D.~A.}\ \bibnamefont {Drabold}}, \bibinfo {author} {\bibfnamefont {M.~P.}\ \bibnamefont {Grumbach}}, \ and\ \bibinfo {author} {\bibfnamefont {R.~M.}\ \bibnamefont {Martin}},\ }\bibfield  {title} {Unconstrained minimization approach for electronic computations that scales linearly with system size,\ }\href {\doibase 10.1103/PhysRevB.48.14646} {\bibfield  {journal} {\bibinfo  {journal} {Physical Review B}\ }\textbf {\bibinfo {volume} {48}},\ \bibinfo {pages} {14646} (\bibinfo {year} {1993})}\BibitemShut {NoStop}%
\bibitem [{\citenamefont {Yang}(1997)}]{Yang1997}%
  \BibitemOpen
  \bibfield  {author} {\bibinfo {author} {\bibfnamefont {W.}~\bibnamefont {Yang}},\ }\bibfield  {title} {Absolute-energy-minimum principles for linear-scaling electronic-structure calculations,\ }\href {\doibase 10.1103/PhysRevB.56.9294} {\bibfield  {journal} {\bibinfo  {journal} {Physical Review B}\ }\textbf {\bibinfo {volume} {56}},\ \bibinfo {pages} {9294} (\bibinfo {year} {1997})}\BibitemShut {NoStop}%
\end{thebibliography}
\end{document}